# Growth Patterns of Subway/Metro Systems Tracked by Degree Correlation

Daniel E. Whitney
MIT Engineering Systems Division
77 Massachusetts Ave
Cambridge MA 02139
e-mail: dwhitney@mit.edu

ABSTRACT

Urban transportation systems grow over time as city populations grow and move and their transportation needs evolve. Typical network growth models, such as preferential attachment, grow the network node by node whereas rail and metro systems grow by adding entire lines with all their nodes. The objective of this paper is to see if any canonical regular network forms such as stars or grids capture the growth patterns of urban metro systems for which we have historical data in terms of old maps. Data from these maps reveal that the systems' Pearson degree correlation grows increasingly from initially negative values toward positive values over time and in some cases becomes decidedly positive. We have derived closed form expressions for degree correlation and clustering coefficient for a variety of canonical forms that might be similar to metro systems. Of all those examined, only a few types patterned after a wide area network (WAN) with a "core-periphery" structure show similar positive-trending degree correlation as network size increases. This suggests that large metro systems either are designed or evolve into the equivalent of message carriers that seek to balance travel between arbitrary node-destination pairs with avoidance of congestion in the central regions of the network.

Keywords: metro, subway, urban transport networks, degree correlation

## Motivation

Newman [1] identified the degree correlation $r$ of a network as q distinctive metric capable of distinguishing network types according to whether it was positive or negative. Although subsequent work [2] showed by example that $r$ cannot reliably separate networks by type, it is nevertheless agreed that networks with positive or negative degree correlation have qualitatively different appearance. Stars have negative degree correlation while grids have positive. In this paper, we focus on urban metro systems and try to discover what their degree correlation or degree correlation history can tell us about them. The long term goal, like that of network theory in general, is to relate form to function.

Young metro systems are small and, not surprisingly, small metro systems are star-like and have negative $r$. As these systems grow, many (but not all) add lines that do not follow the star pattern but instead build up a central core that seems grid-like[1], retaining or adding rays outward to serve suburban populations with increasing size, wealth, and social influence. Each city is different, and some, due to geography, retain the essentially star-like pattern (New York, Boston, for example). But most large systems (London, Tokyo, Shanghai, Seoul, for example) grow increasingly grid-like. In addition, several systems have circles in addition to central cores and radial arms.

One can imagine various mechanisms for describing this growth. Possibly the cores develop because many lines are built and these inevitably cross each other. Possibly they develop to serve and link multiple commercial centers that emerge over time (in Tokyo think of Shinjuku, Shimbashi, Ueno, etc., each large enough to be a city in itself).

In spite of the particular circumstances of each city, those whose metro systems have a central core all exhibit very small negative or decidedly positive degree correlation, and over time their degree correlation, starting quite negative, has grown more positive. This invites an investigation into common reasons or enablers of this commonly observed characteristic. The approach taken here is to imagine that metro systems are "like" canonical forms whose growth patterns we can model analytically in terms of increasing numbers of nodes and whose degree correlation we can calculate explicitly. Are they "like" trees with cross-links? Are they regular grids with tails sticking out? Are they a set of rays with one or more concentric circles? And what if they are? Can we learn anything from this?

## Literature

Many authors have applied network science methods to transportation systems of all kinds, including airlines, roads, city streets, and urban transport (bus, trolley, subway, etc.). [3][4][5][6][7][8][9][10][11] A general review of spatial networks appears in [12]. Some authors model every metro station as a node [3] while others model only terminal and line transfer nodes. [8] Some make bipartite models to separate lines from transfers.[3] Some papers seek power law behavior, a difficult effort [8] since these networks generally have too few nodes to allow statistically confident determination of power law behavior. Others seek robustness of these systems to node removal. A comprehensive analysis [3] observes few consistencies in these networks, a reflection of the different geographical and other constraints that apply to them. Interesting characterizations of metro system structure and function are given in [9]. The metrics in this paper called degree connectivity and complexity permit growth phases to be identified, corresponding to the evolution of degree correlation identified here.

Network Science has inherited a number of metrics from social network theory, such as average degree, clustering coefficient, and degree correlation. [13] Social network researchers abstracted social networks but were always able to link their metrics to actual entities; nodes were usually people while edges were relationships between those people.

[1] We use terms like "grid," "star," and "core-periphery" loosely for illustrative purposes. Actual metro systems are difficult to characterize precisely in words.

Network Science has sought to understand networks at a consistently high level of abstraction, often suppressing entirely the actual nature of the nodes and edges in the hope of obtaining generalizable insights. Numerous successes have resulted, but often something is lost when the context and its constraints are omitted.

The Pearson degree correlation $r$ is usually calculated numerically for real or simulated networks. Theoretically it is zero for E-R random networks. It is not usually calculated for regular networks because these do not present an ensemble in the spirit of typical statistical analyses. The same limitation might be associated with the clustering coefficient. Nevertheless, closed form expressions are known for the clustering coefficient of regular networks as well as for the Watts-Strogatz Small World network. [1] Closed form expressions for average path length and network diameter have been calculated for many regular networks such as rectangular grids, stars, stars with concentric circles, and so on. [14] The purpose of these calculations is not fundamentally statistical but rather to gain some structural insight. Closed form expressions for degree correlation and clustering coefficient of some regular networks are derived here for the same purpose.

## Organization of the Paper

The paper is organized as follows. Data on a number of large metro systems are presented first, followed by historical growth data on a few systems for which old maps are available. Then we analyze a number of regular networks to see if any of them can be made to fit the historical patterns and from this to see if we can infer some relationships between structure and function. Appendices provide information on analysis methods as well as a summary of all the formulae derived.

## Urban Rail and Subway Networks

We begin by looking at the data. Most urban metro systems provide websites for users that contain maps of the contemporary systems, and many provide historical information. In addition, subway fans maintain their own websites or Wikipedia pages with maps, photos, and historical data including old maps. So we have a rich trove of data from which to work, although data from fans do not have official vetting and could contain errors.[2] Since some of the historical data needed are available nowhere else, we rely on these unofficial sources anyway. For each metro system, we construct a network model on which we calculate various metrics, discussed below.

### Modeling Assumptions

Our modeling method uses nodes to represent transfer stations and terminals. We do not include any intermediate stations. If we did, the degree correlation would be biased

[2] Even old official maps contain occasional errors or ambiguities. Historical events such as abandonment or renaming of stations and the split of the Berlin system from 1961 to 1989 present additional modeling challenges.

toward positive in a non-repeatable way by nodes of degree 2 linked to each other. (This is evident from Table 4 in [3].) Many metro systems grow by extending existing rays and this would add more degree-2 nodes without changing the basic structure. Including only transfer stations and terminals allows us to capture the structure without the bias. [8] Researchers interested in other aspects of these networks include the intermediate stations and, sometimes, their geographic locations, allowing them to study growth patterns such as suburban extension compared to core growth and calculate various efficiency metrics.[15][16]

**General Observations**

As shown below, metro systems begin small and star- or tree-like with negative degree correlation and grow to become grid-like and display positive degree correlation. The older ones were built after the first commuter rail and interurban rail systems had been built around their respective cities. The rail lines terminated at large stations separated some distance from the city center to minimize the effect of coal smoke on the center. Surface trolley lines and underground lines were built to link these train stations, and in most large cities (London, Berlin, Moscow, Tokyo, Madrid, Beijing, to name a few) the subway system includes a "circle line" linking these stations.[3] As these cities grew, other subway lines were added. These additional lines usually are linear rather than circular and cut through the circle as well as each other.

More recently built systems, such as Seoul and Shanghai, emerged in an era when rail travel is less frequently used, so rail terminals do not play as big a role in the siting of stations or the routing of lines. The circles instead provide a generic short-cut opportunity that is especially useful in star-like systems where the alternative would be that everyone rides to the center if their destination is on another line, resulting in longer journeys and congestion at the center. In Tokyo the old Yamanote circle line links the main rail stations but the new Oedo circle line does not.

**Metro System "Missions"**

Metro systems do not grow randomly but seek to serve the transport needs of urban and (more recently) suburban populations. Transportation system designers usually begin with demand data comprising an origin-destination matrix. Generalizing this idea, we speak here loosely of such systems having a higher level function which we call the "mission." A goal of this paper, similar to network research in general, is to see if we can relate the structures and growth patterns we see with a mission, or evolution of missions over time. For some systems, such as Paris, there is a record of the mission(s) over time that drove the design of the system, as discussed below. In other cases we can get clues about a mission-driven structure. Our final observation, that the mission of big metro systems seems to be to act like message routers, is probably not one that occurs naturally to the designers of these systems, but it has appeal to network researchers who seek commonality at a level of abstraction that combines metro systems with the Internet, WANs, and so on.

---

[3] Tokyo has both an elevated circle line (Yamanote) and a subway circle line (Oedo).

The mission of small star/tree systems is to bring people from the periphery to the city center. The vast majority of metro systems are like this today.

Gastner and Newman [17], using a growth algorithm, studied the structure of distributive networks in which there was a central source or sink node. This structure corresponds well to the commuter rail system around Boston, which they studied. It is primarily a star pattern, and has approximately zero clustering coefficient and very negative degree correlation. The mission of this system, heavily influenced by the geography of rivers that flow to the harbor, is to bring suburban residents to the center of the city (North and South Stations), from which they can often walk to nearby business and government centers. The structure of all systems that have a single source or single sink, such as water and sewer, can be easily related to this mission. Star-like metro systems can suffer from congestion at the center (ask anyone who has been in Part Street in Boston during rush hour[4]) but water and sewer systems benefit from this congestion in the efficiency of processing that it affords.

Another pattern that emerges is that of circle lines that join the main rail terminals. London, Moscow, Tokyo, Nagoya, Beijing, Shanghai, and many others, have such circles. Thus we can infer a mission for these systems, namely to allow voyagers arriving by rail to get to destinations inside the city or to transfer from one rail station to another.

Another known mission, implemented in Moscow and Beijing, is to provide interior lines that permit troops to be moved around to defend the city against land invasion. Beijing's first metro was designed in the 1950s and 60s by Soviet engineers with this purpose, patterned in principle after the Moscow system. [26]

The Paris metro was designed in the late 1890s and built up by the 1930s following this design for the most part. It is the most compact and dense metro system in the world and was designed specifically to serve central city dwellers and discourage suburban dwellers from journeying to the city. [21] There is no circle line, so travelers arriving at, say Gare du Nord and wanting to connect through Gare de Lyon, must change metro lines to do so. Several of the other Gares are connected point to point by different metro lines. As the suburbs grew in wealth and influence, the government responded by building the RER, a primarily radial commuter rail system with a few highly congested transfer stations to the Metro such as Chatalêt and the Gares. The RER provides additional point to point connections between some of the Gares. But the RER also contains no circle line. The circle function is provided by highways. As businesses moved to the suburbs, travelers in the outer rings have been forced into cars, and the primary mode for travelers not originating and ending their trips inside the central city is automobile.[18]

---

[4] The author has lived in, and extensively used the metro and commuter rail systems of, Boston, London, Paris, Munich, Tokyo, and Zürich.

**Data and Maps [19][20][21][22][23][24][25][26][27][28][29][30][31][32] [33]**

Table 1 shows present-day statistics for some urban regional rail and subway systems. Figure 1 shows the data numerically. Both the table and the figure are sorted by increasing degree correlation.

The regional rail systems of Boston, Paris, Tokyo and Munich have negative degree correlations. Boston and Munich are like stars while Paris and Tokyo are more grid-like with stubs. The regional rail system of Moscow consists of radial lines plus two circles and has a positive degree correlation. The Moscow subway comprises crossing radial lines and a circle, as does London. Their respective degree correlations are positive. This reflects their denser more grid-like structures. The average nodal degree of these rail and metro systems rarely exceeds 3. Clustering coefficients are usually less than 0.15. Typical interchange stations do not often have more than 2 lines intersecting, except in famously complex stations like Otemachi in Tokyo, which has 5. Both cognitive and physical limitations are clearly involved, as many researchers have observed. Except for degree correlation, only weak trends can be observed among these metrics. The same can be said for the Estrada communicability metric [34], not shown. Meshness ratio [35] measures the extent to which a planar graph achieves the maximum number of connections possible given its number of nodes and edges. Larger meshness ratio is associated with more alternate paths and thus for riders' convenience and network robustness to node or edge deletion. A plot of the meshness ratio $\mu$ vs average nodal degree $< k >$ shows that it almost perfectly fits the theoretical[5] formula $< k > = 4\mu + 2$, indicating (not proving) that metro systems are basically planar. The main exceptions: London has lower than theoretical meshness while Beijing and Nagoya have higher. Tokyo's metro has several non-planar areas.

| Network | Number of nodes | Number of edges | <k> | r | Clust Coeff $c_5$ | Meshness Ratio $\mu$ |
|---|---|---|---|---|---|---|
| Boston Commuter Rail | 26 | 26 | 1.96 | -0.4167 | 0.0256 | 0.0417 |
| Osaka Metro and JR Loop | 40 | 64 | 3.2 | -0.2319 | 0.0942 | 0.3421 |
| Mexico City Metro | 29 | 40 | 2.7586 | -0.2047 | 0.0494 | 0.2407 |
| Milan Metro | 22 | 25 | 2.2727 | -0.1433 | 0.0455 | 0.125 |
| Paris Metro | 83 | 139 | 3.301 | -0.1309 | 0.1619 | 0.358 |
| Boston Metro | 24 | 28 | 2.333 | -0.1063 | 0.075 | 0.1364 |
| Tokyo Regional Rail | 147 | 204 | 2.775 | -0.0911 | 0.0783 | 0.2034 |

[5] Valid for $n >> 1$.

| | | | | | | |
|---|---|---|---|---|---|---|
| Shanghai Metro | 50 | 78 | 2.96 | -0.0765 | 0.0909 | 0.29 |
| Nagoya Metro | 21 | 29 | 2.7619 | -0.0563 | 0.0952 | 0.2632 |
| Paris Commuter Rail | 45 | 50 | 2.222 | -0.047 | 0.0504 | 0.0814 |
| Beijing Metro | 27 | 39 | 2.89 | -0.0461 | 0.1062 | 0.28 |
| Munich Schnellbahn | 50 | 65 | 2.6 | -0.0317 | 0.0892 | 0.177 |
| Seoul Metro | 78 | 122 | 3.128 | -0.0122 | 0.1355 | 0.3026 |
| Moscow Regional Rail | 90 | 121 | 2.6884 | -0.0105 | 0.0515 | 0.1875 |
| New York Subways | 76 | 119 | 3.1316 | 0.0198 | 0.1279 | 0.3041 |
| Tokyo Regional Rail plus Metro | 191 | 300 | 3.1414 | 0.0425 | 0.0897 | 0.2936 |
| Berlin U- and S-bahn | 91 | 152 | 3.34 | 0.0431 | 0.1571 | 0.3539 |
| Madrid Metro | 42 | 73 | 3.4762 | 0.0809 | 0.1621 | 0.4125 |
| Tokyo Metro no Yama | 60 | 105 | 3.5 | 0.13945 | 0.143 | 0.396 |
| London Underground | 109 | 144 | 2.85 | 0.1463 | 0.126 | 0.1729 |
| Moscow Metro | 43 | 68 | 3.163 | 0.1734 | 0.1457 | 0.3171 |
| Barcelona Metro | 67 | 98 | 2.9254 | 0.2197 | 0.0716 | 0.2538 |
| Tokyo Metro + Yama | 68 | 127 | 3.735 | 0.2197 | 0.1375 | 0.454 |
| Moscow Metro plus Regional Rail | 126 | 193 | 3.0635 | 0.3052 | 0.0681 | 0.2782 |

**Table 1. Statistics for Some Urban Rail and Subway Systems at the Present Time, Arranged According to Increasing Degree Correlation. The clustering coefficient is calculated according to Eq (5) in [1]. The notations "No Yama" and "+ Yama" distinguish the Tokyo metro without and with the Yamanote circle line, respectively.**

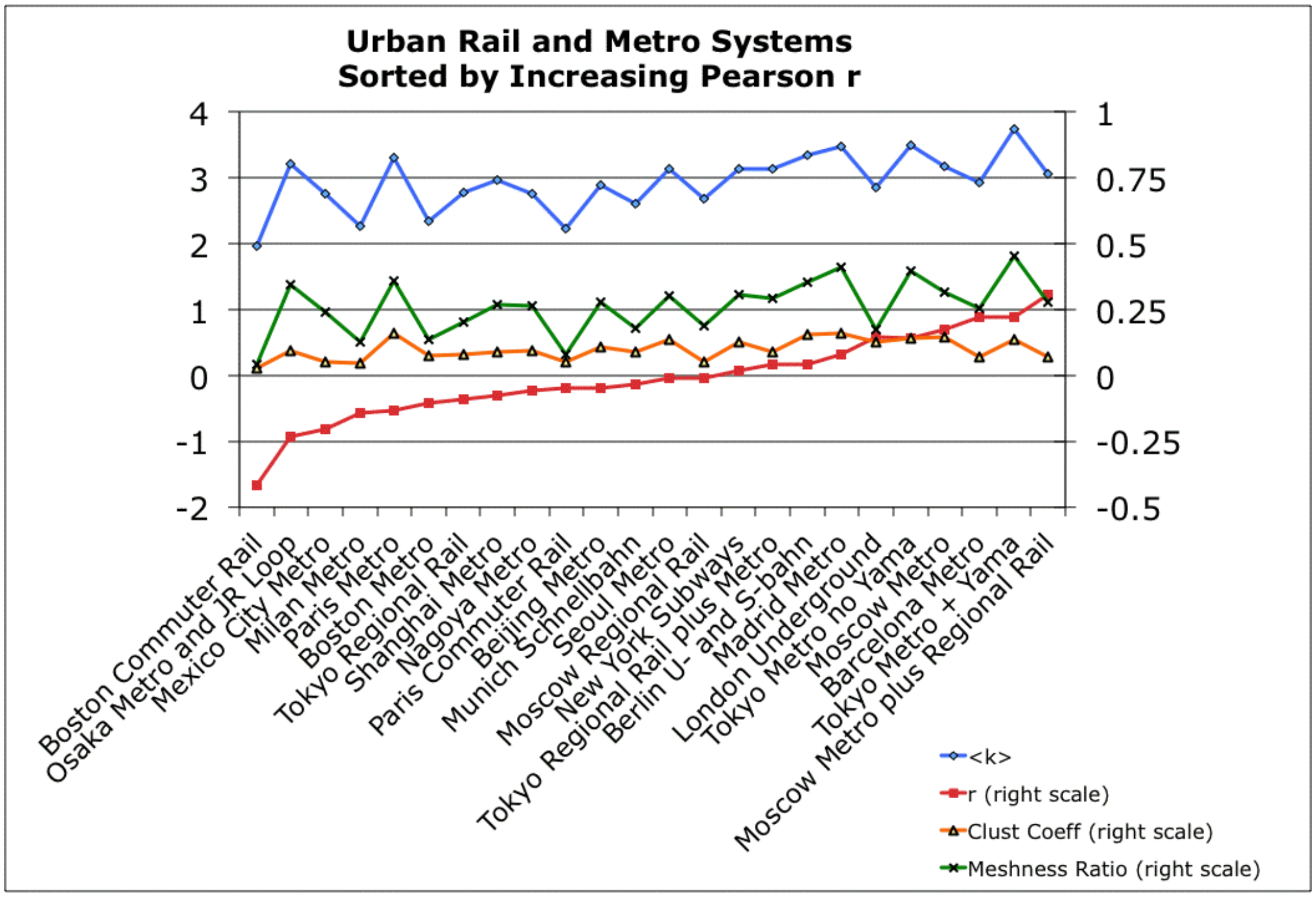

**Figure 1. Data on Some Large Metro Systems**

Figure 1 and Table 1 show that the typical indicators $<k>$, $c_5$, and $\mu$ generally rise as $r$ rises but the relationships are ragged and the changes are not large. Degree correlation shows a strong trend as well as a qualitatively significant sign reversal, indicating that along the spectrum shown a definite change in structure occurs. Large and small systems appear all along the spectrum of $r$ (Figure 2), indicating that increasing $r$ is not associated with increasing network size.

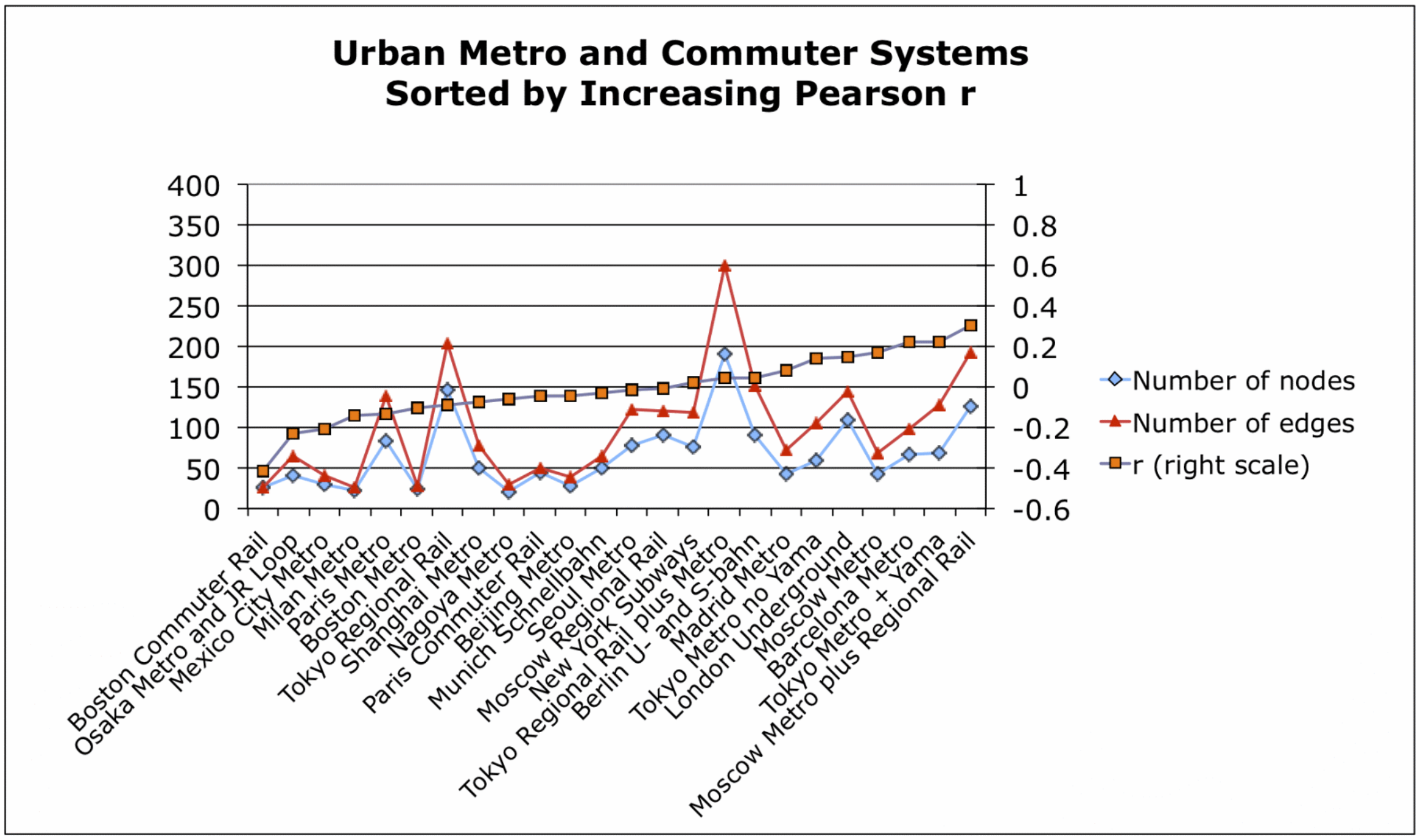

**Figure 2. Comparing Sign of Degree Correlation with Network Size**

**Metro System Histories**

Using old maps it is possible to reconstruct the growth history of some metro systems. The author found, or was able to construct, such histories for the London Underground, and the Moscow, Berlin, Beijing, and Tokyo metros. The growth patterns of these systems reflect or were driven by the growth of the respective cities over many decades. Recent metro systems, such as Shanghai and Seoul, have been built so fast that their emergence cannot be called history in the same sense because the cities did not change as fast as the systems were built.

London's system had a circle line as early as 1889, joining nearly all the main rail stations as it does to this day, plus some branches out into the western suburbs. The system comprised several independently owned lines until 1933, when central planning took over. The history in terms of typical network statistics is shown in Figure 3. Example maps from which the data were obtained are shown in Figure 4. These statistics show that the system has seen its degree correlation rise steadily as new crossing lines have cut the circle and each other, increasing the density of the grid-like structure in the center. This pattern is consistent with a change in service mission for the system. Originally it was set up to link the main rail stations to help travelers coming into the city by rail to get to other rail stations. As more people moved to the suburbs, the need arose for local surface trolley service, which was later placed under ground. Lines serving this need tended to start in one suburb, cross the city, and end in another suburb on the opposite side. Several directions of the compass are now served in this way. The circle now serves additionally to provide a shortcut so that travelers arriving on one radial line need not ride all the way to the center in order to reach a destination on a different radial line.

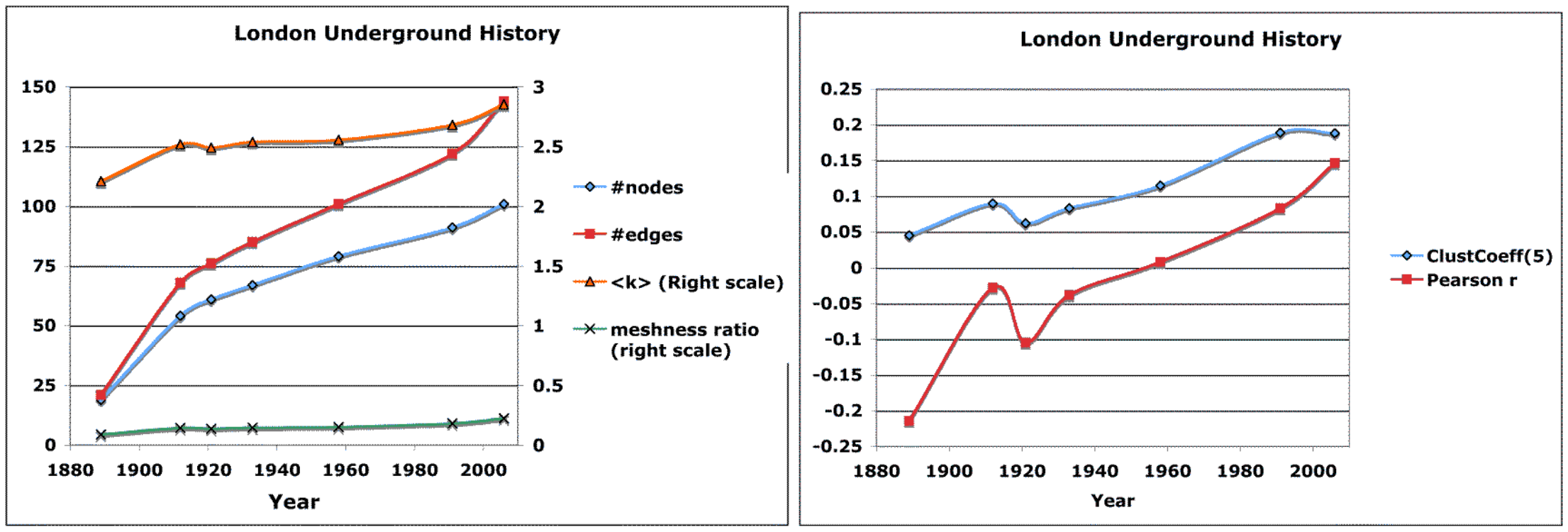

**Figure 3. Growth of the London Underground According to Common Network Statistics. The London Underground has grown steadily in terms of nodes and edges, but hardly at all in terms of average nodal degree. As the original circle shape with branches has been augmented by lines that cross, the network has become a more and more dense grid-like structure, and its degree correlation and clustering coefficient have steadily grown. The degree correlation became positive between 1950 and 1960. Except for 1889, nodes are transfer stations or terminal stations. For 1889, all stations that survive in later maps are the nodes.**

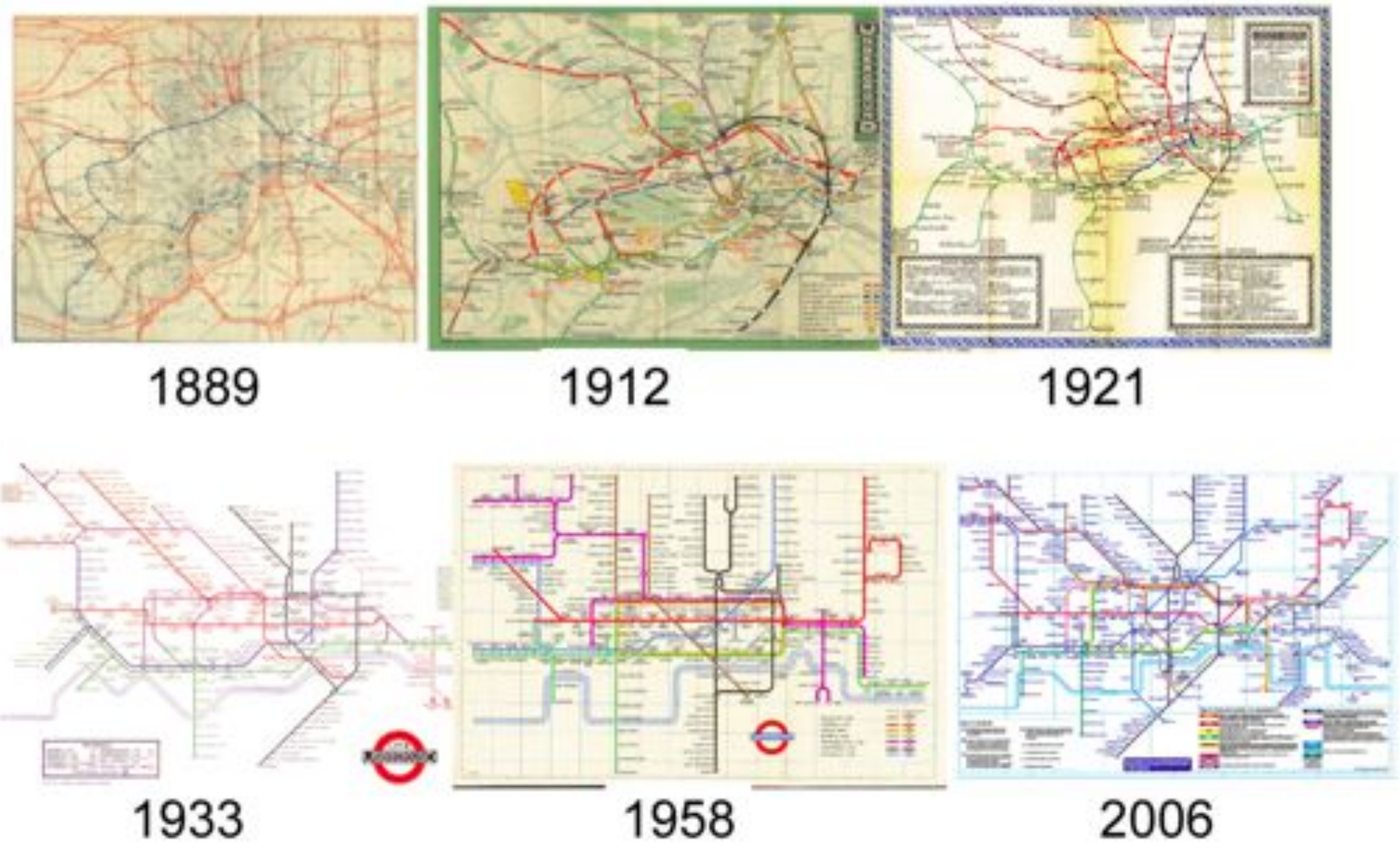

**Figure 4. London Tube Maps for Selected Years from 1889 to 2006. [27]**

The Moscow Metro's history can be documented similarly, using maps dating from the early 1930s. [20] This system was centrally planned from the outset. While the planners of this system intended it to have a circle line joining the city's main rail stations, and maps show the circle before construction began, construction actually occurred between the early 1950s and 1964. Nevertheless, the history of the main network metrics mirrors that of London to a remarkable extent, as shown in Figure 5. Example maps from which the data were obtained are shown in Figure 6.

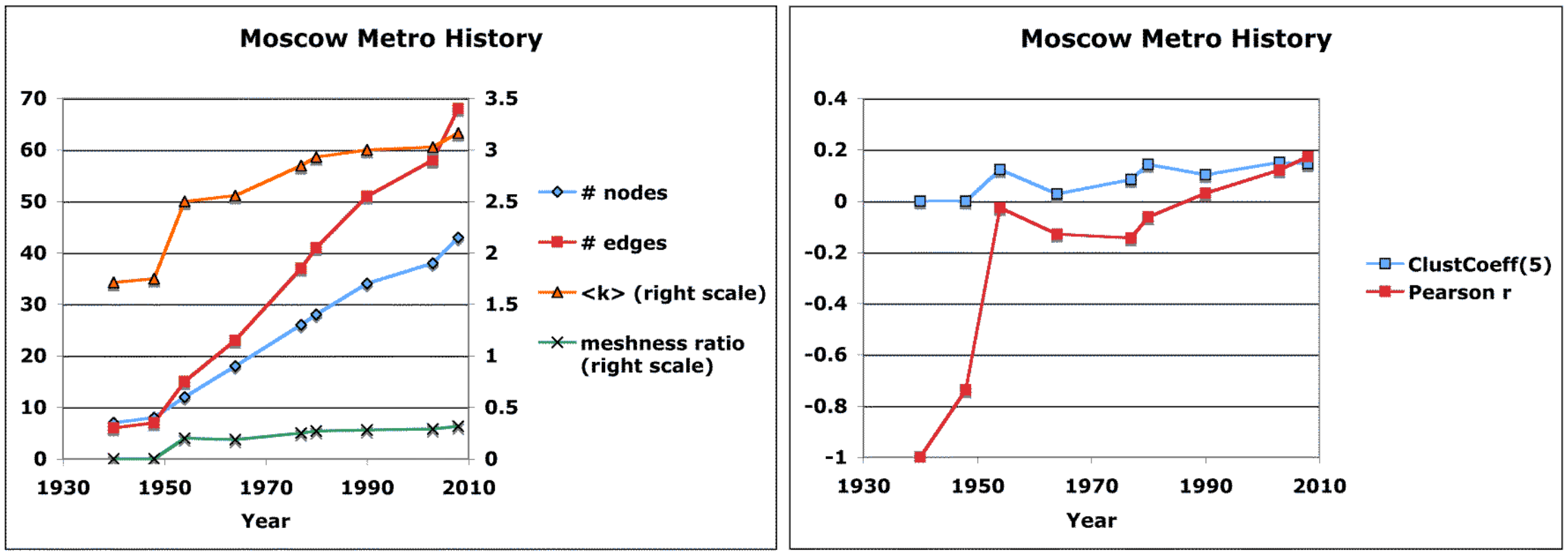

**Figure 5. Growth of the Moscow Metro According to Common Network Statistics. Like London, the Moscow Metro has grown steadily in terms of nodes and edges, but hardly at all in terms of average nodal degree. It is about half the size of the London Underground in terms of nodes and edges but in other respects, such as clustering coefficient, average nodal degree, and degree correlation history, it is quite similar.**

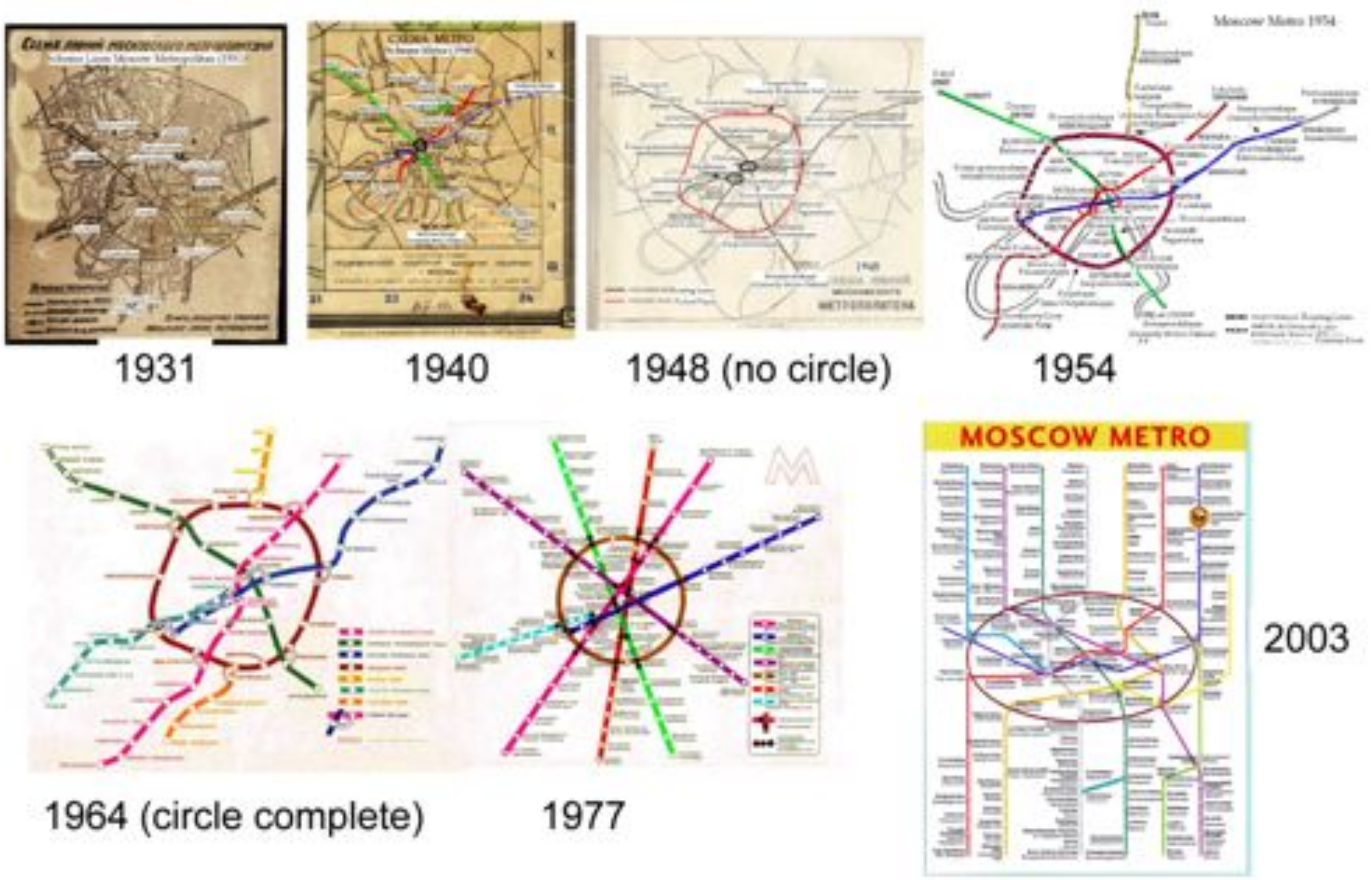

**Figure 6. Moscow Metro Maps for Selected Years. Early maps indicate the existence of a circle but in fact its construction did not begin until the 1950s.**

Figure 7, Figure 8, and Figure 9 show history data for Berlin [24][25], Beijing [26], and Tokyo [30][30] respectively. Beijing's metro was/will be constructed during three distinct periods, the 1950s, the 2000s, and the future out to 2020. These three periods are represented by single data points in Figure 8. The Tokyo history is shown both with and without the Yamanote Line, which predates the subways. The subway circle line (Oedo) began operation in 2000.

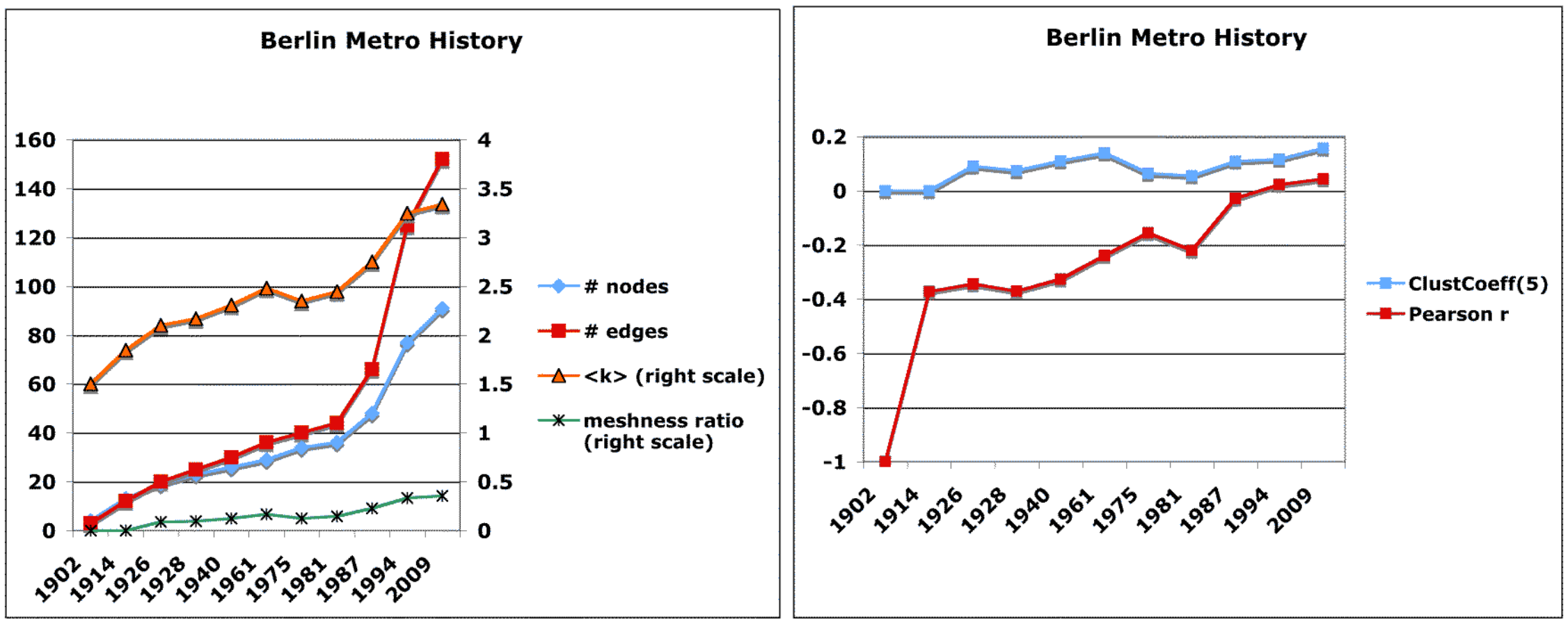

**Figure 7. History of Berlin Metro**

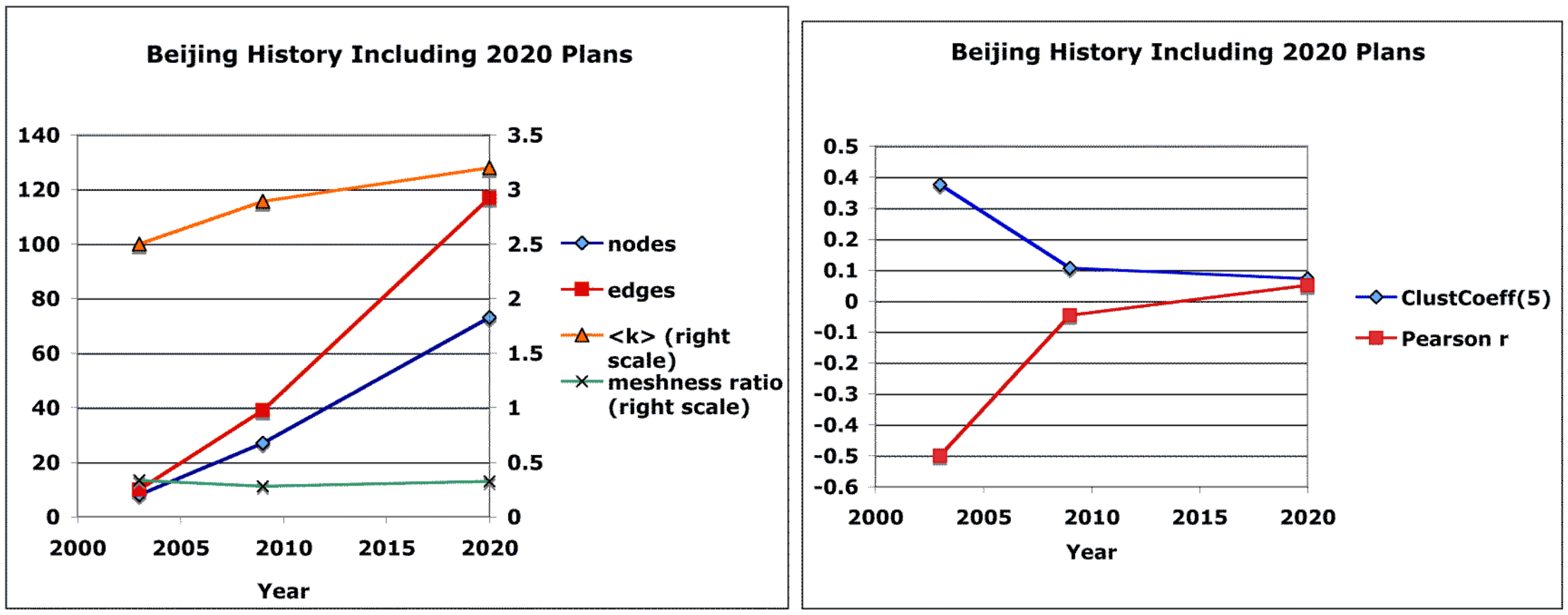

**Figure 8. History of Beijing Metro Including Future Plans for 2020**

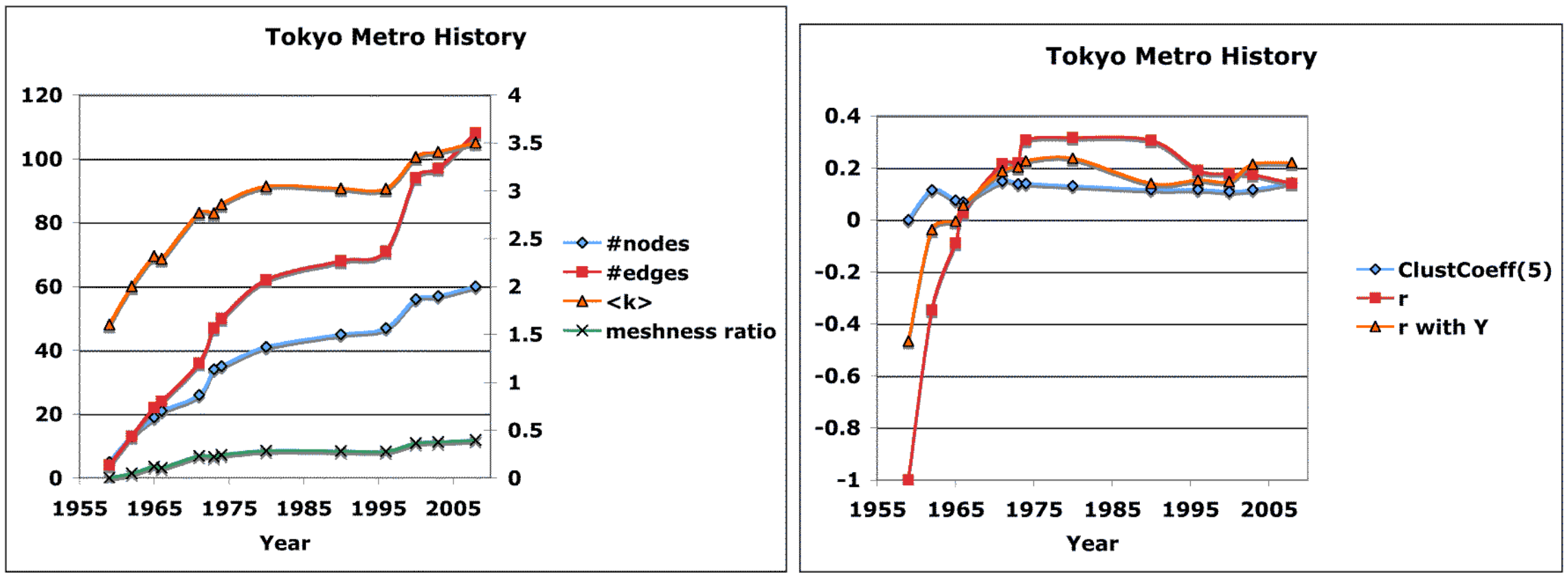

**Figure 9. History of the Tokyo Metro. On the right, the evolution of $r$ is shown with and without the Yamanote Line.**

The common pattern in these histories is a more or less steady rise of the degree correlation from initially quite negative to modestly positive values. At the same time, average nodal degree and meshness ratio grow but stay within narrow limits.

## Regular Network Models

The next step in the analysis is to go beyond the raw data and imprecise terms like "grid" and construct regular graph models that might or might not structurally resemble metro systems. Then we calculate their degree correlation, and in some cases their clustering coefficient, and see if, by growing these regular systems in size, we can reproduce the positive-trending behavior of the metro systems whose histories were discussed in the previous section. Since a number of these regular systems are similar by construction to known types of systems with specific missions, we can see if clues to metro system mission can be inferred. Underlying this approach is the assumption that, if form follows function in networks, then networks with similar forms might have similar functions. Since the link between network form and function is not precise and is still in need of much research, the connections offered here must remain speculative and are offered in the spirit of stimulating ideas.

The regular networks analyzed are listed in Appendix E. Summary of Degree Correlation Formulae for Regular Graphs. The clustering coefficient formulae for the core-periphery models are in Appendix D. Clustering Coefficient for Some Core-Periphery Models. All the models are constructed according to a specific pattern (tree, star, star with circle, square gird, etc.) that resembles at least some aspect of a metro system's structure. In addition, a random growth experiment was performed, described below. For each of these networks, an analytical expression for the degree correlation was derived. The method for doing so is explained in Appendix A. Example Degree Correlation Calculation.

Of all the networks analyzed, only one type (shown in four variations) that resembles metro systems shows increasing degree correlation as network size grows: "HOT," "Abilene," "HOT" plus a circle, and "Abilene" plus a circle.[6] See Figure 10. These networks comprise a partially or fully connected core, a set of "gateway" nodes surrounding the core (linked by circles where present), and some radial lines extending from the gateways. If there are many such radial lines, then the degree correlation is negative, but real metro systems usually have one or rarely two radii extending from each gateway, so the degree correlation is weakly negative or positive. Moscow is one of the clearest examples.

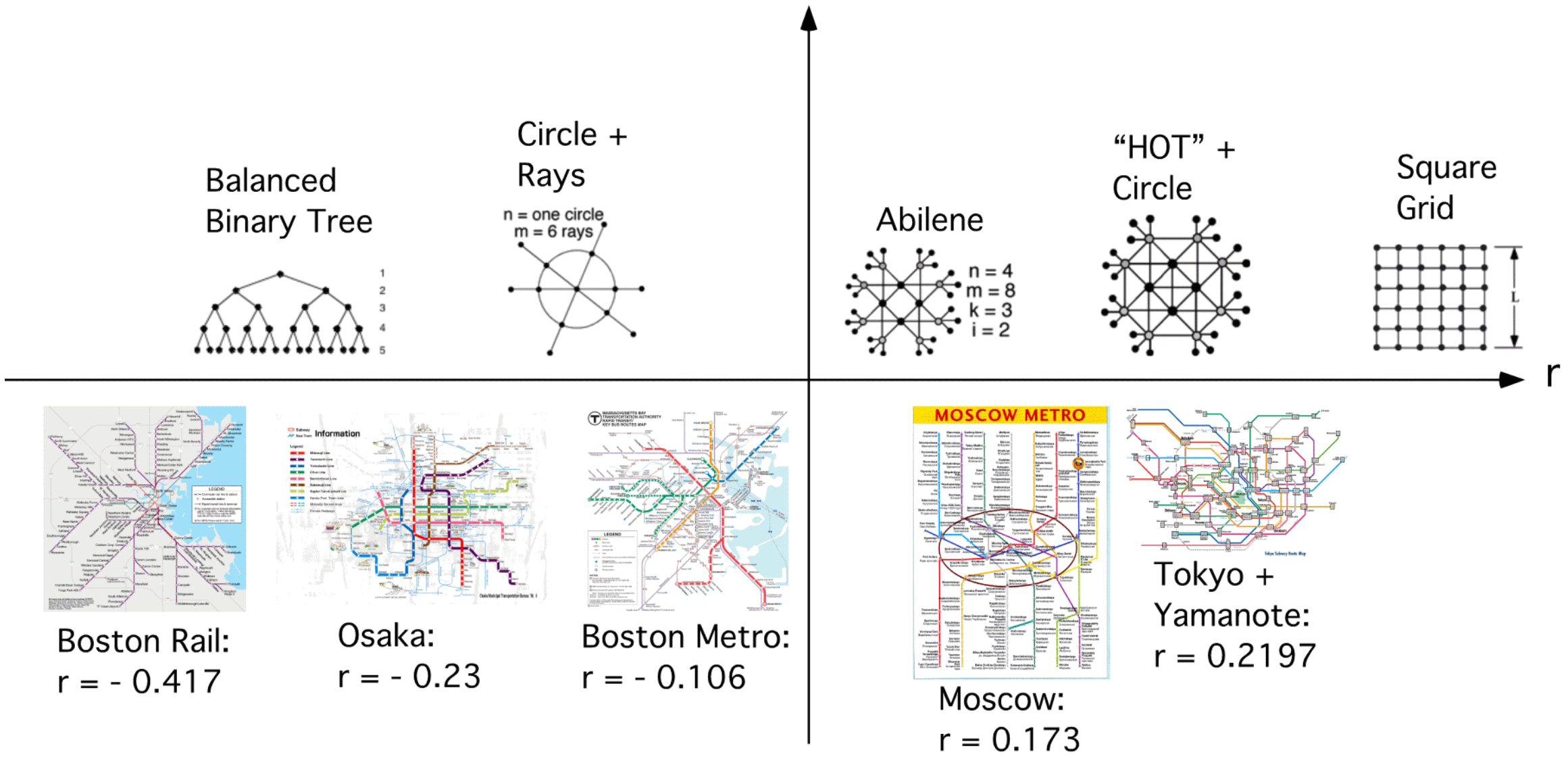

**Figure 10. Illustration of Tendency of $r$ for Various Regular Network Types (above the horizontal axis) Compared to Some Metro Systems (below the horizontal axis).**

[6] These names, as well as the intent behind the designs, are taken from [37].

**The core-periphery types just above $r = 0$ could have $r < 0$ for some parameter choices but will have $r > 0$ for choices that are similar to metro systems.**

Table 2 shows the result of finding the best fit "HOT" plus circle model to each of several large metro systems. The fit was found by numerical search, adjusting the four decision variables $n, m, k,$ and $i$ to match the five descriptor variables $\#\, nodes, \#\, edges, r, c_5,$ and $meshness$. It is unnecessary to match $< k >$ independently. The search finds non-integer values for the decision variables, so the solutions shown in the table represent the nearest integers. The match is reasonably good. It should be kept in mind, however, that this is a functional/numerical match, not a physical configuration match. It does show, however, that the "HOT" plus circle paradigm can achieve descriptive variables quite similar to some real metro systems. The Abilene family cannot generate enough edges per node to allow good matches even though it can achieve positive degree correlation. The other regular models cannot generate positive degree correlation at all.

| | n | m | k | i | #nodes | #edges | <k> | r | c5 | meshness |
|---|---|---|---|---|---|---|---|---|---|---|
| Moscow + Rail HOT+circle | 8 | 31 | 3 | 2 | 132 | 213 | 3.22 | 0.302 | 0.0637 | 0.315 |
| Moscow+Rail actual | | | | | 126 | 193 | 3.06 | 0.305 | 0.068 | 0.278 |
| | | | | | | | | | | |
| Moscow Metro HOT+circle | 5 | 12 | 2 | 2 | 41 | 69 | 3.365 | 0.1658 | 0.127 | 0.371 |
| Moscow Metro actual | | | | | 43 | 68 | 3.16 | 0.173 | 0.1457 | 0.317 |
| | | | | | | | | | | |
| Tokyo no Yama HOT+circle | 9 | 10 | 4 | 2 | 59 | 105 | 3.56 | 0.1376 | 0.1317 | 0.4122 |
| Tokyo no Yama actual | | | | | 60 | 105 | 3.5 | 0.139 | 0.143 | 0.396 |
| | | | | | | | | | | |
| Tokyo+Yama HOT+circle | 11 | 10 | 5 | 2 | 71 | 134 | 3.77 | 0.2056 | 0.136 | 0.4637 |
| Tokyo+Yama actual | | | | | 68 | 127 | 3.735 | 0.2197 | 0.1375 | 0.454 |

**Table 2. Comparison of "HOT" plus Circle Approximation to Some Actual Large Metro Systems. $n$ is the number of core nodes. $m$ is the number of gateway nodes. $k$ is the number of pendant nodes per gateway, while $i$ is the number of core nodes linked to each gateway ($i \le n$).**

In addition to the regular models shown in Figure 10, a random growth experiment was performed. See Appendix B. Estimating Metrics for Randomly Placed Intersecting Lines with a Circle and Appendix C. Random Growth Experiment. Lines were randomly drawn over a circle, imitating a growth process that adds lines instead of nodes. The result is that $r < 0$ consistently. In another random experiment, pendants were added randomly with increasing probability $p_p$ to a square 5x5 grid in order to find a middle ground between a grid with no pendants ($r \to 2/3$) and a grid where every line has a pendant ($r = -1/\text{number of lines}$). In this case we find $r > 0$ until $p_p > 0.2$, indicating that only a small percentage of lines with pendants is needed to drive the degree correlation negative.

## Discussion

The regular networks whose growth results in a degree correlation that trends toward positive are each patterned after wide area networks whose mission is to permit messages to flow efficiently from one peripheral node to another. These regular networks have three kinds of nodes: periphery, gateway, and core (router). This pattern is condensed in our real network modeling scheme into terminal, representing all the stations on a line outside the core, plus the core's transfer stations. Some actual metro systems have clearly identifiable gateway nodes, such as those that occupy the circles of Moscow, Tokyo and London.

While we have not conducted any flow analyses on our regular networks, the literature supports the idea that systems with periphery-gateway-core router structures serve message routing purposes. The literature presented here comprises the original design of AT&T's long distance public switched telephone system (PSTN), the work by Li et al on power laws and degree correlation, and the work of Dodds, Watts, and Sabel on communication efficiency in various kinds of organizational structures.

The AT&T system [36] was designed to have 53 tightly linked regional centers (the routers) distributed around the US, with each regional center fed by one or more primary outlets (the gateways) linked to several toll centers which in turn served a city having several central offices, each with up to several thousand individual subscribers. The shape of this system is core-periphery. The mission was to connect two callers with the minimum number of electrical connections, each of which degraded sound quality. The system was designed top-down in this way but as traffic patterns emerged, short cuts were built between toll centers.

Li et al [37] studied scaling, power laws, and degree correlation. To illustrate their points, they constructed several "toy" networks having up to 1000 nodes and having a power law degree distribution. Of these, one was patterned after the Abilene WAN whose structure is core-periphery like the PSTN while the others were, respectively, one constructed using preferential attachment, one a deliberate "bad design" for communication routing purposes, and one linked randomly. They then showed that the Abilene-like network had the largest message carrying capacity by a wide margin.

Dodds, Watts, and Sable [38] constructed trees and systematically added links in various ways to see the effect on message-carrying efficiency and avoidance of congestion. The best systems for both message-carrying and robustness to node deletion were those containing links among the higher levels of the tree, creating in effect a core-periphery system. (Note in Appendix E. Summary of Degree Correlation Formulae for Regular Graphs that adding cross-links to a balanced binary tree above the leaf level pushes $r$ toward less negative values compared to a tree with no cross links.)

These papers do not prove that metro systems function as passenger routers but the structures are similar and, when the number of leaves is cut to match the few leaves of metro systems, the cited systems have positive degree correlation or trend toward positive as their size increases.

## Conclusions

In this paper we derived closed form expressions for the Pearson degree correlation for some regular networks. Trees, and circles with rays, have highly negative degree correlation while bounded grids have highly positive degree correlation. Intermediate forms have intermediate degree correlation. Networks with core-periphery structure can have positive or negative degree correlation depending on parameter values that govern their structure. Those with structure similar to metro systems have positive degree correlation.

We then found some real metro networks that were respectively tree-like or grid-like and showed that their degree correlations have the predicted sign. Of all the graph and network theory metrics typically tried (average nodal degree, clustering coefficient, etc.), only degree correlation clearly separates these systems structurally and over time, and shows wide variation. We also showed how five metro networks, the London Underground and the Moscow, Berlin, Beijing, and Tokyo Metros, evolved from tree-like to grid-like over their lifetimes by adding crossing linear structures to an original or eventual circle with branches, and tracked this progression as the evolution from negative to positive degree correlation. Randomly evolved structures that are superficially similar to growing metro systems cannot duplicate their history of steadily increasing degree correlation although they are similar in average nodal degree and clustering coefficient. Finally we made the connection between metro systems and wide area networks using regular forms and their analytically-derived degree correlation and inferred that a major mission, intended or emergent, of metro systems is to treat the riders like messages and route them efficiently from their origin to their destination.

## Appendix A. Example Degree Correlation Calculation

The formula for calculating the Pearson degree correlation is

**Equation 1**

$$r = \frac{\sum(x-\bar{x})(y-\bar{y})}{\sqrt{\sum(x-\bar{x})^2\sum(y-\bar{y})^2}}$$

$x$ and $y$ are the nodal degrees of nodes linked to each other. Since every node is represented in each list, $\{x\}$ and $\{y\}$ are the same lists so $\bar{x} = \bar{y}$.

The method for calculating the Pearson degree correlation for the balanced binary tree with no cross-linking is illustrated in Figure 11. It shows the classic Pearson array for a balanced binary tree with 5 layers, with the first layer having $N = 1$. Entries in all layers beyond the second are either $(1-\bar{x})(3-\bar{x})$ or $(3-\bar{x})^2$. For large $N$ the first and second lyers in the array contribute nothing. The third layer is repeated and the repeated sets of entries contribute a total number of rows which, in the limit of large $N$, is the total number of rows in the whole array minus the number of rows in the last two layers. The total number of rows is $ksum$. Since every entry in one column is represented in the other column, $\bar{x} = \bar{y}$.

Equation 2 shows how to calculate $\bar{x}$.

**Equation 2**

$$\text{Sum of row entries} = \sum k_i^2 = 10 * 2^{N-1} - 14$$

$$\text{Total number of rows} = ksum = \sum k_i = 2^{N+1} + 4$$

$$\bar{x} = \frac{\sum k^2}{\sum k} = \frac{\langle k^2 \rangle}{\langle k \rangle} = 2.5 \text{ as } N \to \infty$$

To evaluate the sum in the Pearson numerator we need to know how many entries there are of the form $(1-\bar{x})(3-\bar{x})$ and how many there are of the form $(3-\bar{x})^2$. All of the former appear in the last two layers and there are $2^N$ of them. Thus there are approximately $ksum - 2^N$ of the latter. Thus we obtain for the numerator

**Equation 3** $\text{numerator} \approx 2^N(3-\bar{x})(1-\bar{x}) + (ksum - 2^N)(3-\bar{x})^2$ .

To calculate the denominator we need to sum the square of the entries in one column. In the left column there are $2^{N-1}$ entries of the form $(1-\bar{x})$ and thus there are $ksum - 2^{N-1}$ entries of the form $(3-\bar{x})$. Thus the denominator is

**Equation 4** $\text{denominator} \approx 2^{N-1}(1-\bar{x})^2 + (ksum - 2^{N-1})(3-\bar{x})^2$

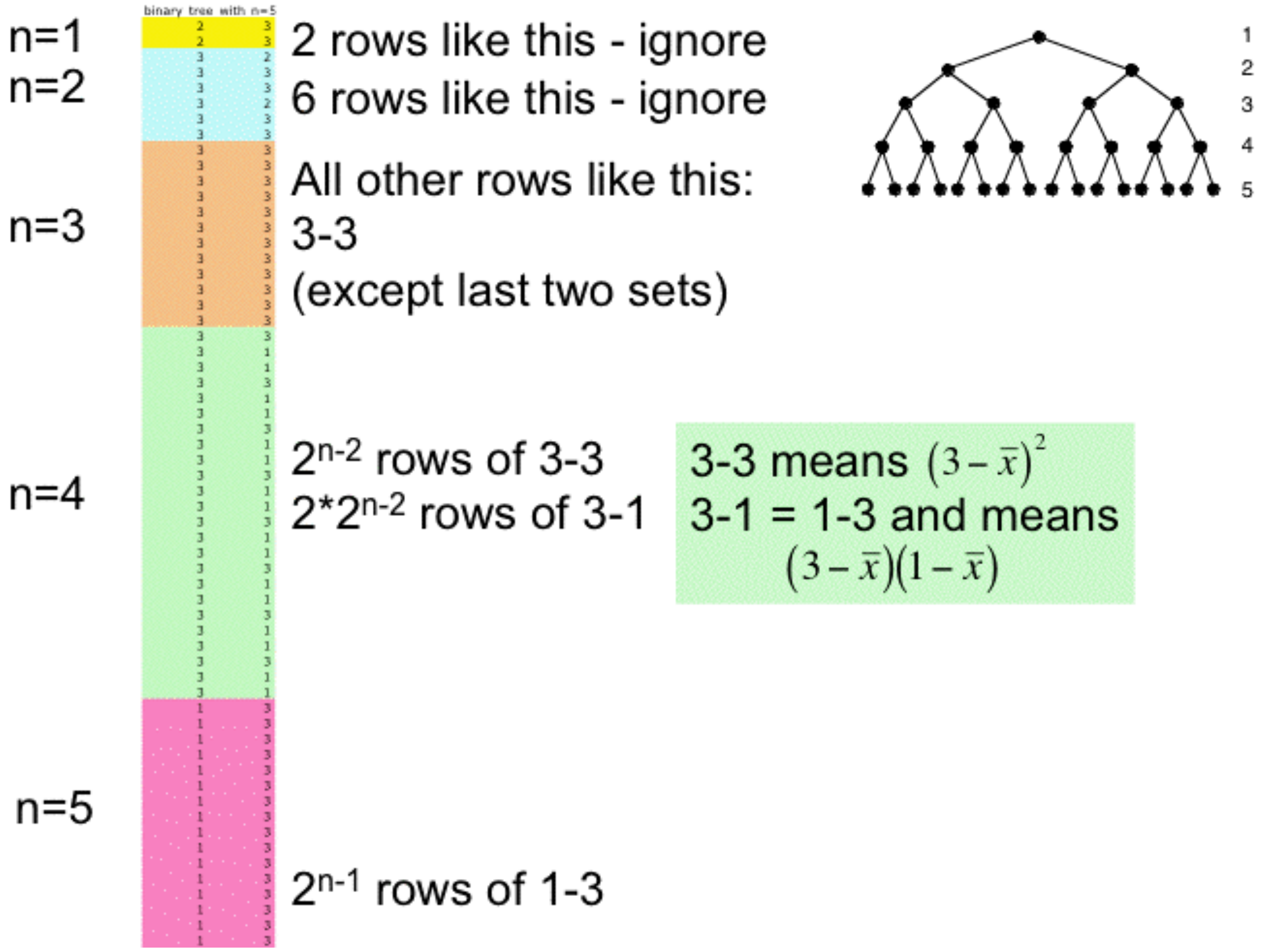

**Figure 11. Sketch of the Calculation of the Numerator of the Pearson Degree Correlation for a Balanced Binary Tree with 5 Levels. The Pearson array appears at the left. A census of entries of various types of entries in the numerator appears to the right of the array. For this tree, $r = -0.4122$ .**

The result of these calculations is shown in Figure 12.

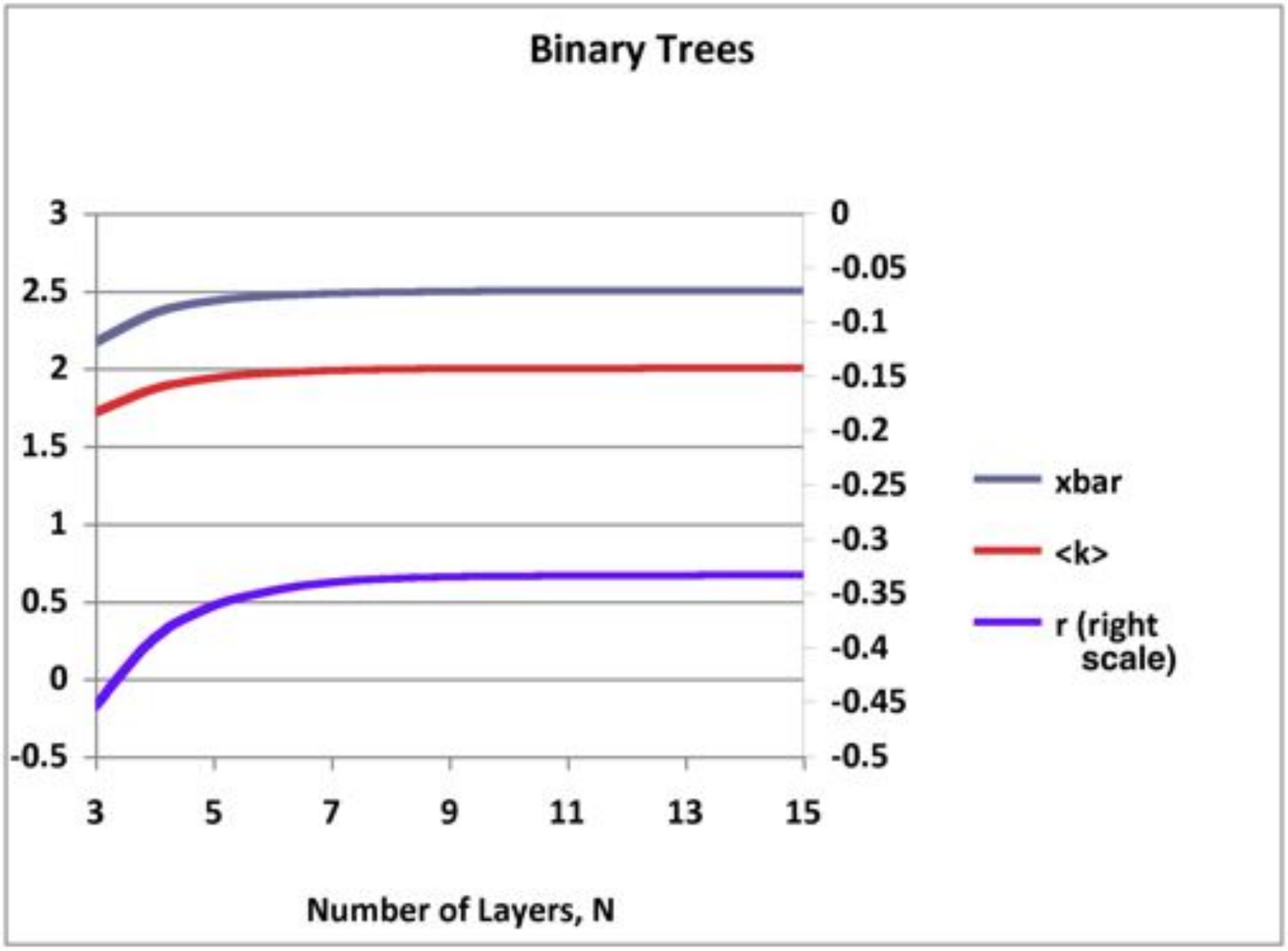

**Figure 12. Pearson Degree Correlation for Binary Trees**

## Appendix B. Estimating Metrics for Randomly Placed Intersecting Lines with a Circle

Intersecting lines form a planar graph. If there are $L$ lines plus the circle, and if we terminate each line with a node, then we will have at most $L(L-1)/2$ intersections between lines and $2L$ intersections between the lines and the circle. The simulated networks do not have as many intersections as anticipated by the theory because some of the lines are arbitrarily terminated near where they exit the circle so as to more nearly replicate the structure of the metro systems. In addition, the author took the liberty of merging nodes that were very near each other, occasionally removing or adding triangles. Real metro systems are altered in similar ways to create new transfer stations where none existed before, especially if the system has been consolidated by a central planning authority from disparate independent networks. Thus we assume that the number of intersections is multiplied by a factor $\alpha$.[7]

This gives rise to $\alpha L(L-1)/2$ nodes of degree = 4 plus $2L$ intersections between lines and the circle (again giving rise to that many nodes of degree = 4) and $2L$ stub terminals

[7] Simulations show that $\alpha \sim 0.5$ is a good estimate.

(giving rise to $2L$ nodes of degree = 1), respectively, for a total of at most $\left(\alpha L^2 + (8-\alpha)L\right)/2$ nodes and an average nodal degree of

**Equation 5** $$z = \frac{4\alpha L(L-1)/2 + 4(2L) + 2L}{\alpha L(L-1)/2 + 4L}$$

If $L = 10$ and $\alpha = 0.5$ we have an estimated 62.5 nodes and average nodal degree of 3.04.

**A. Degree Correlation**

To calculate $r$, we note that the $2L$ end stub nodes of degree 1 link to $2L$ nodes of degree 4 where lines cut the circle. Each of these circle nodes has three links to nodes of degree 4 (two of these adjacent on the circle) and one link to a stub node of degree 1. The $\alpha L(L-1)/2$ interior nodes of degree 4 have links to other nodes of degree 4.

Then the number of rows in the Pearson array is

$$num_rows = 2L + 2L + 3 \times 2L + \alpha L(L-1)/2.$$

The sum of the entries in one column is

$$sum_row_entries = 2L \times 1 + 2L \times 3 + 2L \times 3 \times 4 + 4 \times 4\alpha L(L-1)/2.$$

Then

$$\bar{x} = \frac{sum_row_entries}{num_rows}$$

The numerator of the Pearson calculation is

$$num = 2L(1-\bar{x})(4-\bar{x}) + 2L(1-\bar{x})(4-\bar{x}) + 2L \times 3(4-\bar{x})^2 + 4\alpha L(L-1)/2\,(4-\bar{x})^2$$

The denominator is

$$den = 2L(1-\bar{x})^2 + 2L(4-\bar{x})^2 + 3 \times 2L(4-\bar{x})^2 + 4\alpha L(L-1)/2\,(4-\bar{x})^2$$

Then

**Equation 6** $$r = \frac{4L(1-\bar{x})(4-\bar{x}) + 6L(4-\bar{x})^2 + 2\alpha L(L-1)(4-\bar{x})^2}{2L(1-\bar{x})^2 + 8L(4-\bar{x})^2 + 2\alpha L(L-1)(4-\bar{x})^2}$$

If $L = 10$ and $\alpha = 0.5$ then $r = -0.1176$.

**B. Clustering Coefficient**

For the following discussion, we use the definition of clustering coefficient per node given by Newman [1] Eq 5:

**Equation 7**
$$c_{i\,Eq\,5}(k) = \text{number of triangles on vertex } i \Big/ \binom{k}{2}$$

According to Solomon [41] Chapter 3, page 44, randomly intersecting lines on a plane form a Poisson field of random lines. The majority of intersections in such a graph are nodes of degree = 4. The likelihood that a polygon in such a structure is a triangle is ~0.36, a quadrilateral ~0.38, a pentagon ~0.19, and so on in rapidly decreasing likelihood (ibid p 54). Any node not a member of a triangle will have clustering coefficient = 0. Interior nodes, mostly of degree = 4, will thus touch 4 polygons. The likelihood that a polygon is a triangle is $p_T = \sim 0.36$. Let $N_T$ be the number of triangles touching a node of degree $k$. We know that $0 \le N_T \le k$. Then the clustering coefficient of such a node is

**Equation 8**
$$c_{Eq\,5} = \sum_{N_T=0}^{k} \frac{\binom{k}{N_T} N_T p_T^{N_T} (1-p_T)^{k-N_T}}{\binom{k}{2}}$$

Using $p_T = 0.36$ and $k = 4$ yields $c_{Eq\,5} = 0.24$. Nodes on the circle will be connected to half as many polygons as interior nodes since one of their neighbors is a stub. These $2L$ nodes will have half the clustering coefficient of interior nodes. The remaining $2L$ stubs will have zero clustering coefficient. Thus the network's clustering coefficient will be approximately

**Equation 9**
$$c_{Eq5} = \frac{0.24\alpha L(L-1)/2 + 0.12(2L)}{\alpha L(L-1)/2 + 4L}$$

For a network comprising 10 lines and a circle enclosing all the interior nodes, we will have $45\alpha$ interior nodes of degree = 4 plus 20 nodes on the circle of degree = 4 plus 20 terminal nodes of degree = 1. The average clustering coefficient will then be 0.1248. These results plus three random simulations are compared in Figure 14.

## Appendix C. Random Growth Experiment

A growth experiment was performed to see if the growth processes exhibited by London and Moscow (lines added to a circle) could be reproduced in a context-free way. A circle was drawn on a piece of paper and 10 randomly oriented lines were drawn. These lines were grouped into "epochs" composed of the first 4 lines, the next 3, and the last three, simulating the growth of a network like London or Moscow. This kind of growth model is more appropriate than node-based models since metro systems grow by adding whole lines that usually cross the entire existing system and terminate at single terminal nodes. This model better matches Moscow than London, which grew this way at first but later expanded more by branching and extending existing lines rather than by adding entirely new ones.

The typical network metrics were calculated for each of the epochs. An example result appears in Figure 13. The experiment was repeated a total of three times, with results summarized in Figure 14. The average nodal degree, degree correlation, and node clustering coefficient (Equation 5 in [1]) for such networks were also calculated using Equation 5, Equation 6 and Equation 9.

The average nodal degree and clustering coefficient have values and follow histories similar to London and Moscow but the degree correlation follows a history better represented by a circle that accumulates rays. Two explanations are offered. First, thinking purely structurally based on the metrics, we observe that the real networks more resemble grids than a circle with rays. Thinking functionally, we note that the real networks are not constructed randomly. They follow urban growth, and new lines are added in such a way as to provide coordinated access to other lines via important interchange stations. These interchanges (Oxford Circus, Piccadilly, etc.) are well-known for being the most heavily used as well as being frequent destinations. [5][6] Thus, while the randomly grown networks have histories of average nodal degree and clustering coefficient similar to the real ones, their more structurally significant degree correlation history bears no relation to that of the real ones, regardless of whether the real ones are centrally planned or not.

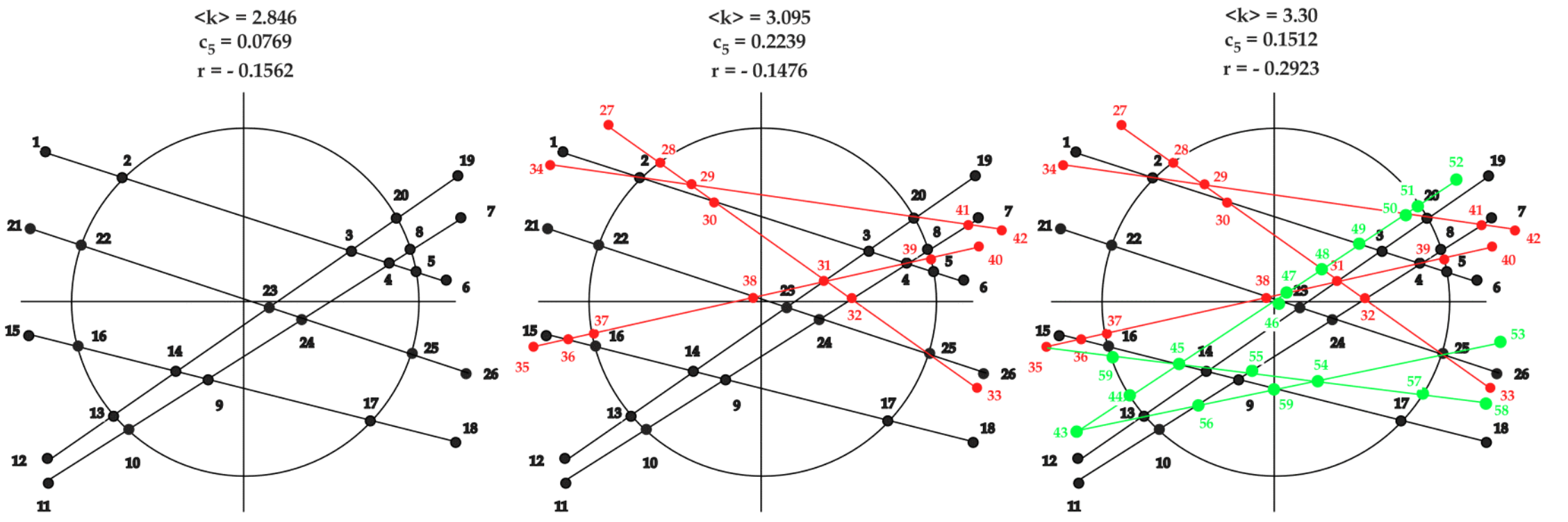

**Figure 13. Evolution of a Randomly Constructed Metro System with a Circle Line and Other Lines Crossing. Three epochs are shown, each with its own network metrics.**

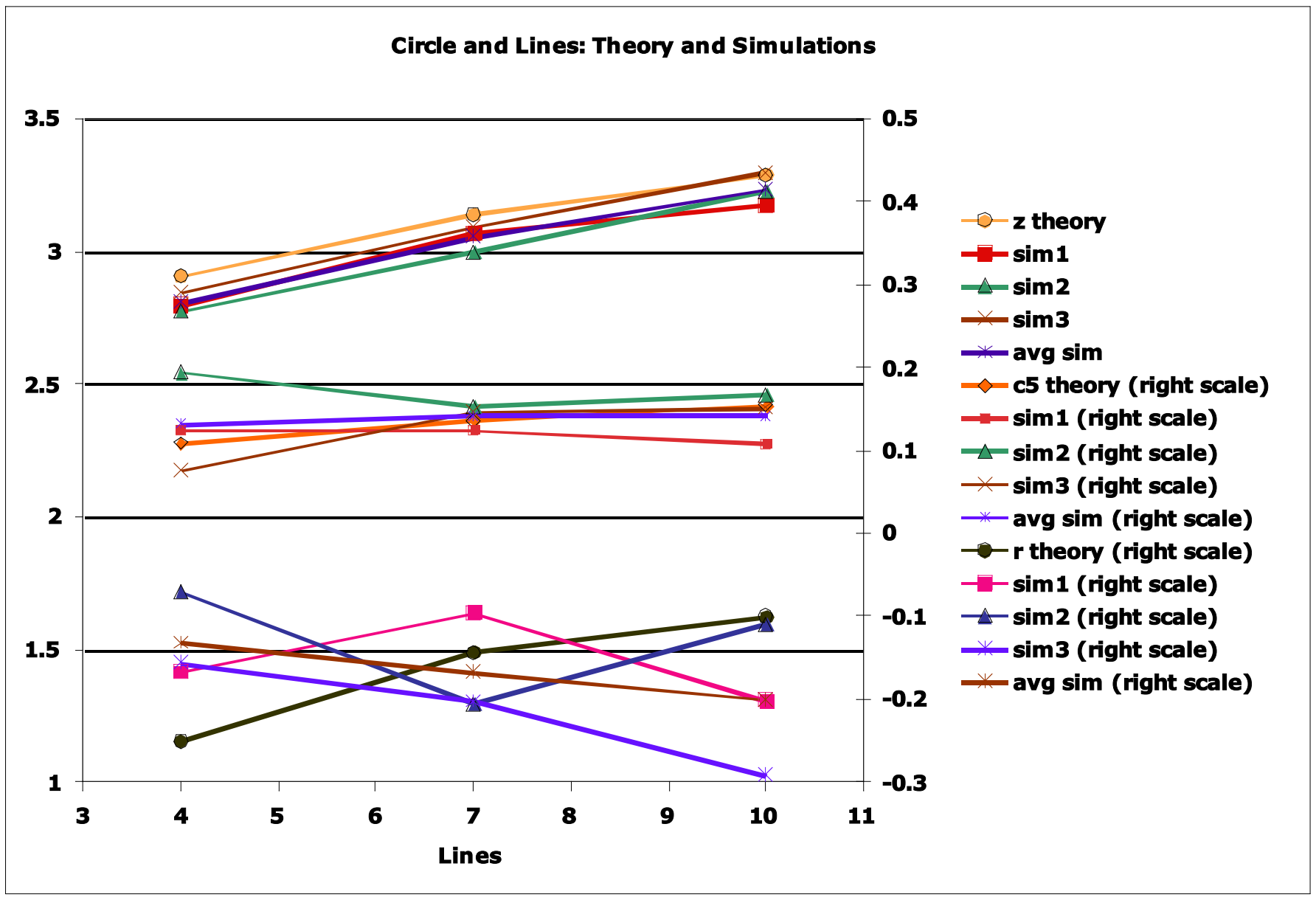

**Figure 14. Results of Simulating a Circle with Lines Crossing at Random.**

## Appendix D. Clustering Coefficient for Some Core-Periphery Models

### 1. HOT model

- n core nodes
- m intermediate nodes
- k pendants of each intermediate node

i = number of core nodes linked to by each intermediate node

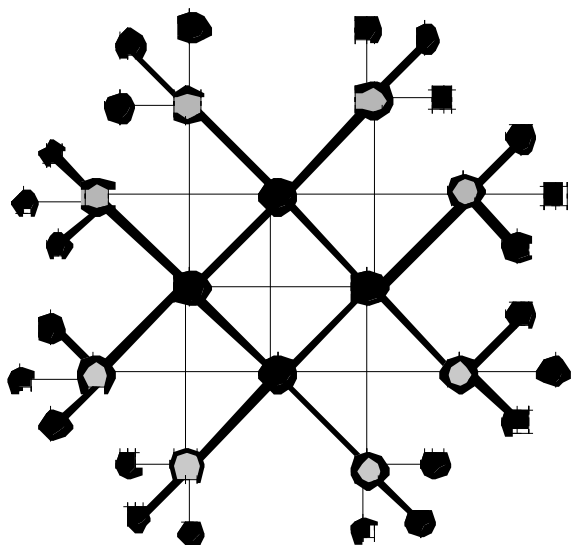

Using Watts-Strogatz (Newman Eq 5) clustering coefficient

$$c_{i\,Eq\,5}(k) = \text{number of triangles on vertex } i \Big/ \binom{k}{2} = \frac{2 * \text{number of triangles on vertex } i}{k(k-1)}$$

For $n$ core nodes: each is in $mi/n$ triangles (if $i > 1$) associated with $i$ intermediate nodes plus $(n-1)(n-2)/2$ triangles with each other. Each core node has degree $d_{core} = n - 1 + mi/n$.

For $m$ intermediate nodes: each is in $i-1$ triangles associated with core nodes. Each intermediate node has degree $d_{\text{int}} = i + k$.

The pendant nodes are not in any triangles.

There are $n + m + mk$ nodes altogether.

Then we have

$$c_{Eq\,5} = \frac{\dfrac{2n\left(mi(i>1)/n + (n-1)(n-2)/2\right)}{(n-1+mi/n)(n-2+mi/n)} + \dfrac{2m(i-1)}{(i+k)(i+k-1)}}{n+m+mk}$$

For $n = 4, m = 8, i = 2, k = 3$, we get $c_{Eq\,5} = 0.059259$ . For $n = 4, m = 4, i = 1, k = 3$, we get $c_{Eq\,5} = 0.1$.

**2. Abilene model**

- ● n core nodes
- ○ m intermediate nodes
- ● k pendants of each intermediate node

i = number of core nodes linked to by each intermediate node

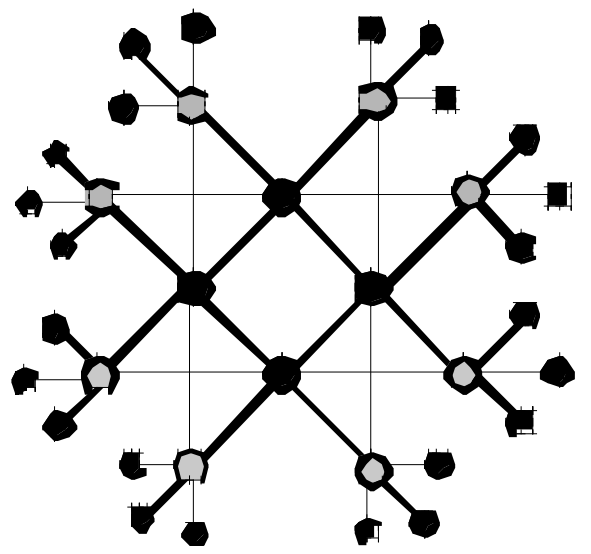

For $n$ core nodes: each is in $mi/n$ triangles (if $i > 1$) associated with $i$ intermediate nodes. Each core node has degree $d_{core} = n - 1 + mi/n$.

For $m$ intermediate nodes: each is in $i-1$ triangles associated with core nodes. Each intermediate node has degree $d_{\text{int}} = i + k$.

The pendant nodes are not in any triangles.

There are $n+m+mk$ nodes altogether.

Then we have

$$c_{Eq\,5} = \frac{\dfrac{2n(mi(i>1)/n)}{(n-1+mi/n)(n-2+mi/n)} + \dfrac{2m(i-1)}{(i+k)(i+k-1)}}{n+m+mk}$$

For $n=4, m=8, i=2, k=3$, we have $c_{Eq\,5}=0.05185$ while for $n=4, m=4, i=1, k=3$, we have $c_{Eq\,5}=0$.

**3. Abilene plus circle**

- n core nodes
- m intermediate nodes
- k pendants of each intermediate node

i = number of core nodes linked to by each intermediate node

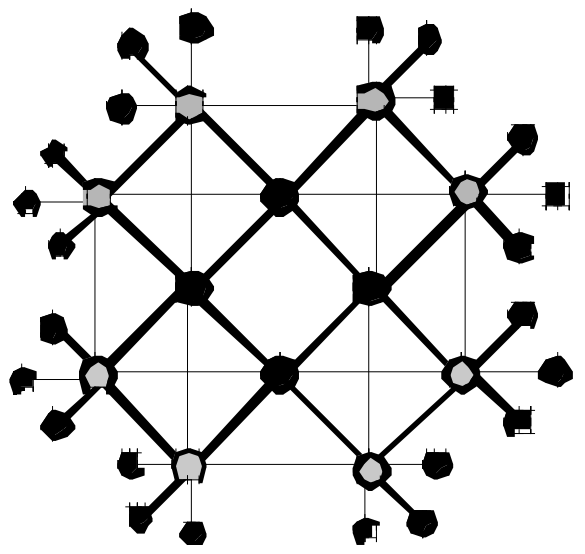

Each of $n$ core nodes is in $mi(i>1)/n+1$ triangles with $m$ internal nodes. Their degree is $mi/n+2$. Each of $m$ intermediate nodes is in $(i-1)+3$ triangles. Their degree is $k+i+2$. Then we have

$$c_{Eq\,5} = \frac{\dfrac{2n[mi(i>1)/n+3]}{(mi/n+2)(mi/n+1)} + \dfrac{2m[(i-1)+3]}{(k+i+2)(k+i+1)}}{n+m+mk}$$

For $n=4, m=8, i=2, k=3$, we have $c_{Eq\,5}=0.094179$ .

**4. HOT plus circle**

- n core nodes
- m intermediate nodes
- k pendants of each intermediate node

i = number of core nodes linked to by each intermediate node

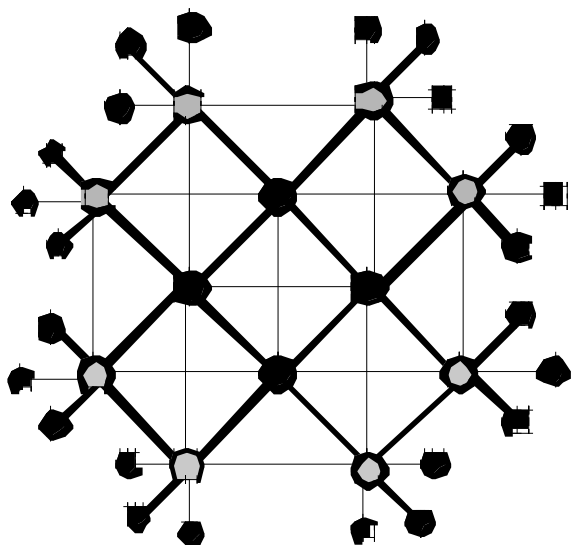

Each of $n$ core nodes is in $mi(i>1)/n+3+(n-1)(n-2)/2$ triangles. Their degree is $mi/n+n-1$. Each of $m$ intermediate nodes is in $(i-1)+1+2$ triangles. Their degree is $k+i+2$.

The total number of triangles is $n[mi(i>1)/n+3+(n-1)(n-2)/2]+m(i+2)$. The total number of nodes is $n+m+mk$.

Then we have

$$c_{Eq\,5}=\frac{\dfrac{2n[mi(i>1)/n+3+(n-1)(n-2)/2]}{(mi/n+n-1)(mi/n+n-2)}+\dfrac{2m(i+2)}{(k+i+2)(k+1+1)}}{n+m+mk}$$

For $n=4, m=8, i=2, k=3$, we have $c_{Eq\,5}=0.14421$

## Appendix E. Summary of Degree Correlation Formulae for Regular Graphs

| Network | Drawing | $\bar{x}$ | Numerator | Denominator | Comments |
|---|---|---|---|---|---|
| Balanced binary tree with $N$ layers | 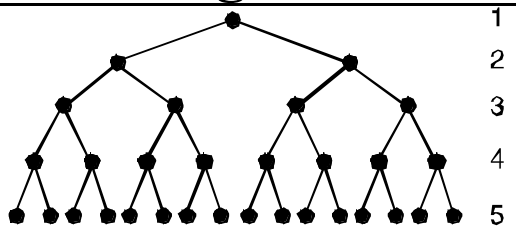  | $\dfrac{10*2^{N-1}-14}{2^{N+1}+4}$ | $\sim 2^{N}(3-\bar{x})(1-\bar{x})+$<br>$(ksum-2^{N})(3-\bar{x})^2$ | $\sim 2^{N-1}(1-\bar{x})^2+$<br>$(ksum-2^{N-1})(3-\bar{x})^2$ | $r \to -1/3$ rapidly<br>$as\ N \to \infty$<br>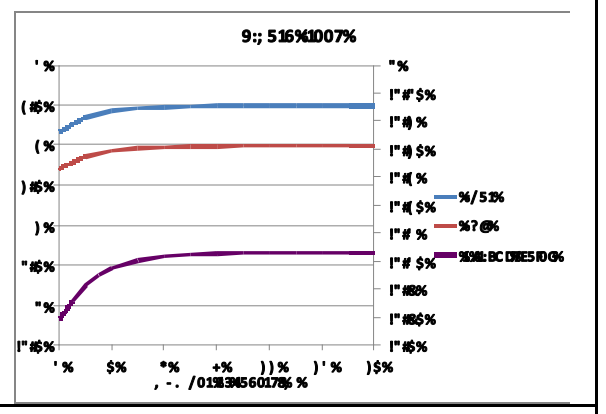 |
| BBT with cross-links | 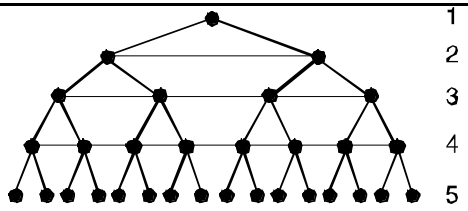  | $\dfrac{13*2^{N}-64}{3*2^{N}-10}$ | $\sim 2^{N}(5-\bar{x})(1-\bar{x})+$<br>$(ksum-2^{N})(5-\bar{x})^2$ | $\sim 2^{N-1}(1-\bar{x})^2+$<br>$(ksum-2^{N-1})(5-\bar{x})^2$ | $r \to -1/5$ rapidly<br>$as\ N \to \infty$ |
| Tree with branching ratio $b$ and $N$ layers | 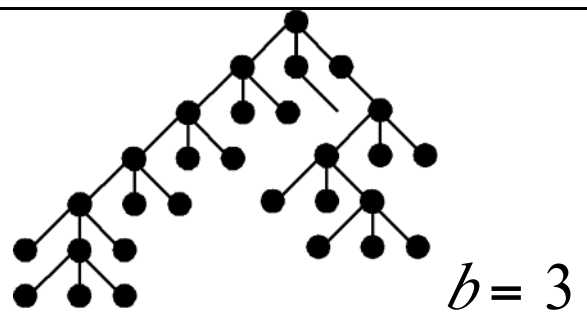  | $(((b+1)^2+1)*b^{(N-1)}-3*b^2-b)/$<br>$[(b*(b-1)+(b+1)*b^{(N-1)}$<br>$-b*(b+1)+(b-1)*b^{(N-1)})]$ | $2*b^{N-1}*(b+1-\bar{x})*(1-\bar{x})+$<br>$(ksum-2*b^{N-1})*(b+1-\bar{x})^2$ | $b^{N-1}*(1-\bar{x})^2+$<br>$(ksum-b^{N-1})*(b+1-\bar{x})^2$ | $r \to -1$ rapidly<br>$as\ b \to \infty$<br>*almost insensitive to* $N$<br>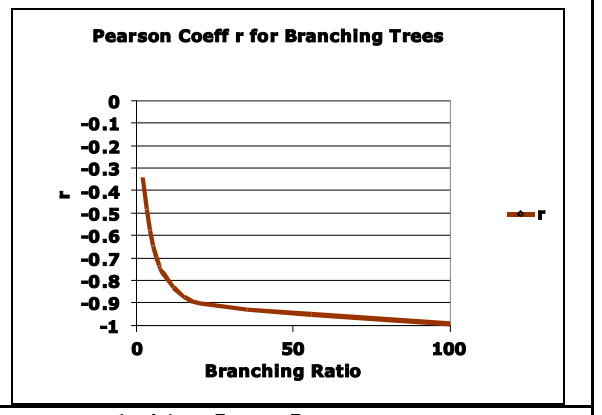  |
| Square grid with $L$ rows and columns | 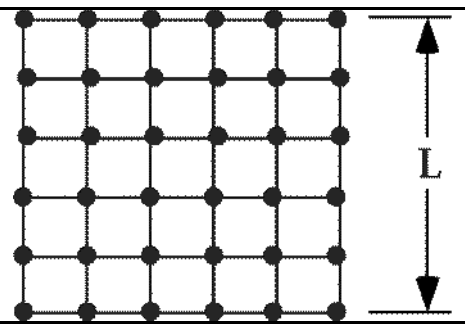  | $4-\dfrac{3a+4}{a^2+3a+2}$<br>where $a=L-2$ | $16(2-\bar{x})(3-\bar{x})$<br>$+8(L-3)(3-\bar{x})^2$ | $8(2-\bar{x})^2$<br>$+12(L-2)(3-\bar{x})^2$ | $r \to 2/3$ slowly<br>$as\ L \to \infty$ |

 Macintosh HD:Users:danielwhitney:Documents:pearson stuff:degree correlation paper:metropaper:Whitney_metro_growth_patterns_021312.doc

<table>
<tr><td></td><td></td><td></td><td></td><td></td><td>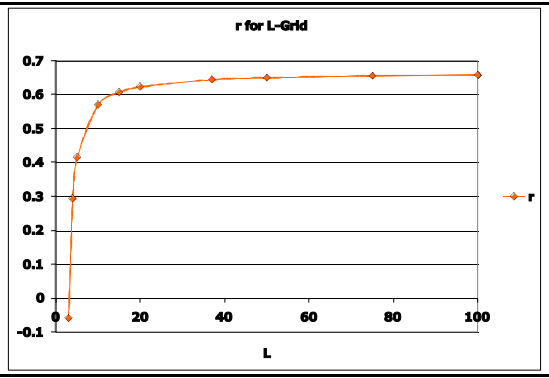</td></tr>
<tr><td>Square grid with $L$ rows and columns and $4L$ pendants</td><td>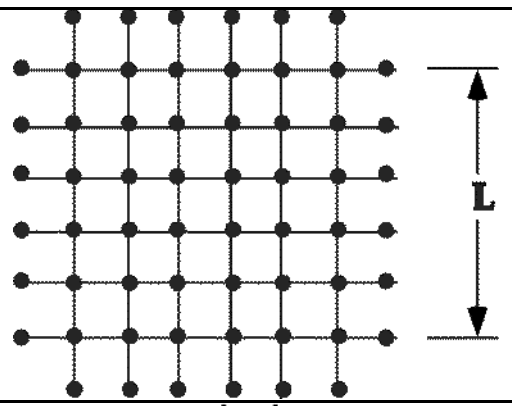</td><td>$\frac{4L+1}{L+1}$</td><td></td><td></td><td>$r=-\frac{1}{L}$ for all $L>0$</td></tr>
<tr><td>Star with $n$ rays and $m$ circles</td><td>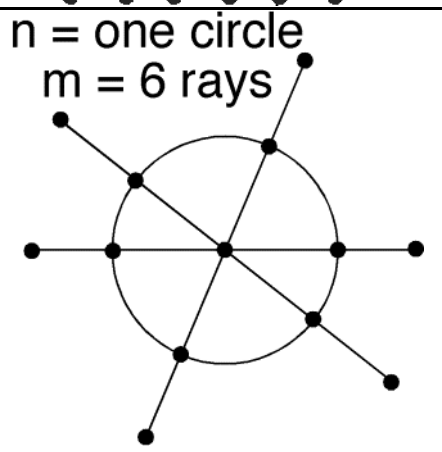</td><td>$\frac{n+16m+1}{4m+2}$</td><td>$2n\left[\begin{array}{l}(n-\bar{x})(4-\bar{x})+2m(4-\bar{x})^2\\+2(4-\bar{x})(1-\bar{x})\\+(m-1)(4-\bar{x})^2\ (m>1)\end{array}\right]$</td><td>$n\left[\begin{array}{l}(n-\bar{x})^2+(2m+1)(4-\bar{x})^2\\+(1-\bar{x})^2+\\2(m-1)(4-\bar{x})^2\ (m>1)\end{array}\right]$</td><td>$r<0$ unless $n\sim 7$<br>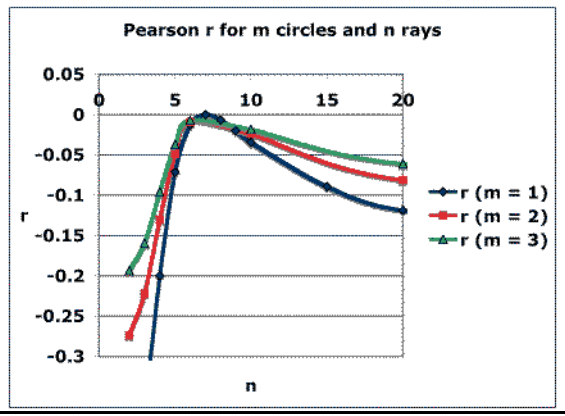</td></tr>
<tr><td>“HOT” with $n$ core nodes all linked to each other, $m$ intermediate nodes, $k$ pendants per intermediate node, and $i$ links from an intermediate node to each core node</td><td>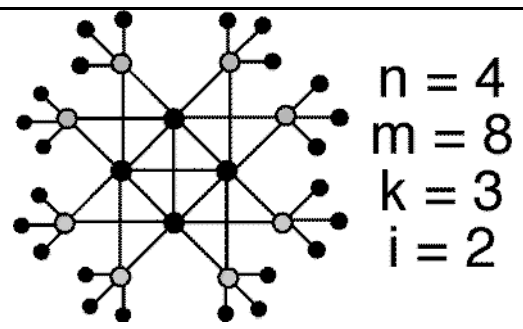<br>● n core nodes<br>○ m gateway nodes<br>• k pendants of each gateway node<br>i = number of core nodes linked to by each gateway node<br>1 ≤ i ≤ n</td><td>$\#rows=n(n-1+mi/n)$<br>$+m(i+k)+mk$<br>$row_sum=n(n-1+mi/n)^2$<br>$+m(i+k)^2+mk$<br>$\bar{x}=row_sum/\#rows$</td><td>$(n-1+mi/n-\bar{x})^2\,n(n-1)$<br>$+2(n-1+mi/n-\bar{x})(i+k-\bar{x})mi$<br>$+2(i+k-\bar{x})(1-\bar{x})mk$</td><td>$(n-1+mi/n-\bar{x})^2$<br>$*\,n(n-1+mi/n)$<br>$+(i+k-\bar{x})^2\,m(i+k)$<br>$+(1-\bar{x})^2\,mk$</td><td>$r<0$ if $k$ is large and $n$ is small<br>or<br>$r>0$ if $n$ is large and $k$ is small<br>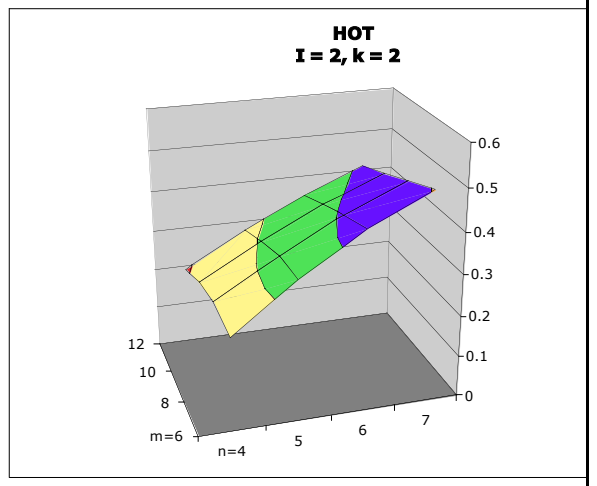<br>$i=2,\ k=2$</td></tr>
</table>

| | | | | | |
|---|---|---|---|---|---|
| "Abilene" like "HOT" except that core nodes link only to two immediate neighbors. | 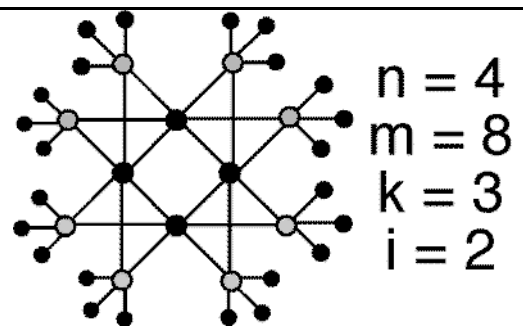 | $\# rows = n\left(2 + mi / n\right)$<br>$+m\left(i + k\right) + mk$<br>$row_sum = n\left(2 + mi / n\right)^2$<br>$+m\left(i + k\right)^2 + mk$<br>$\bar{x} = \# rows / row_sum$ | $2n\left(mi / n + 2 - \bar{x}\right)^2$<br>$+2mi\left(mi / n + 2 - \bar{x}\right)\left(i + k - \bar{x}\right)$<br>$+2mk\left(i + k - \bar{x}\right)\left(1 - \bar{x}\right)$ | $\left(mi / n - 2 - \bar{x}\right)^2 n\left(mi / n + 2\right)$<br>$+\left(i + k - \bar{x}\right)^2 m\left(i + k\right)$<br>$+\left(1 - \bar{x}\right)^2 mk$ | $r < 0$ *unless*<br>$m / n$ *and* $k / i$ *are both small*<br>$i = 2,\ k = 2$<br>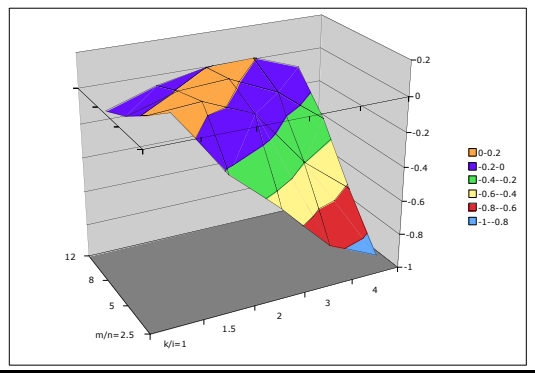 |
| "HOT" as above plus a circle linking the intermediate nodes | 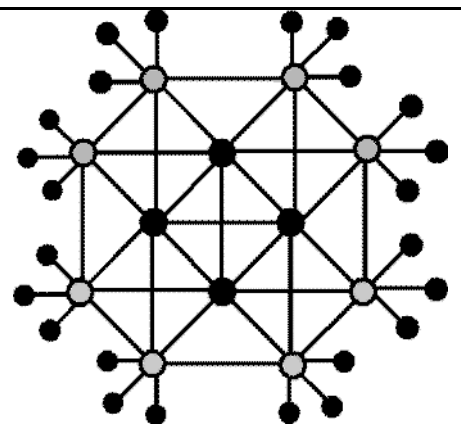 | $\# rows = n(n - 1 + mi / n)$<br>$+m(i + k + 2) + mk$<br>$row_sum =$<br>$n(n - 1 + mi / n)^2$<br>$+m(i + k + 2)^2 + mk$<br>$\bar{x} = \# rows / row_sum$ | $(n - 1 + mi / n - \bar{x})^2 n(n - 1)$<br>$+2(n - 1 + mi / n - \bar{x})$<br>$* mi(i + k + 2 - \bar{x})$<br>$+2m(i + k + 2 - \bar{x})^2$<br>$+2km(i + k + 2 - \bar{x})(1 - \bar{x})$ | $(n - 1 + mi / n - \bar{x})^2$<br>$* n(n - 1 + mi / n)$<br>$+(i + k + 2 - \bar{x})^2$<br>$* m(i + k + 2)$<br>$+mk(1 - \bar{x})^2$ | $r < 0$ especially if $k$ is large<br>*or*<br>$r > 0$ if $n$ is large<br>$i = 2,\ k = 2$<br>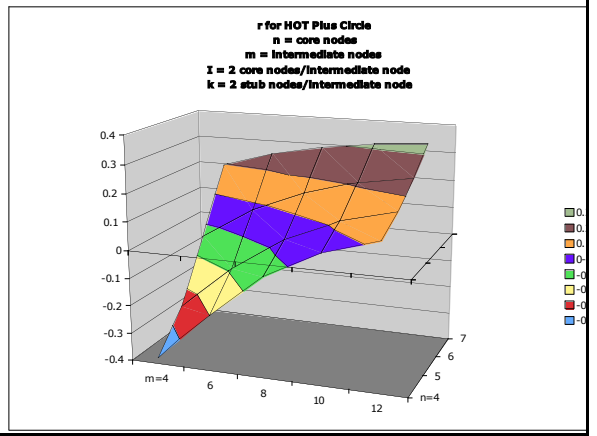 |
| "Abilene" as above plus a circle linking the intermediate nodes | 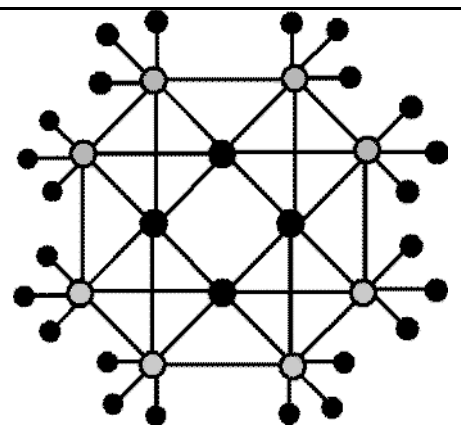 | $\# rows = n(mi / n + 2)$<br>$+m(i + k + 2) + mk$<br>$row_sum =$<br>$n(mi / n + 2)^2$<br>$+m(i + k + 2)^2 + mk$<br>$\bar{x} = \# rows / row_sum$ | $2n(mI / n + 2 - \bar{x})^2$<br>$+2mI(mI / n + 2 - \bar{x})(k + I + 2 - \bar{x})$<br>$+2m(I + k + 2 - \bar{x})$<br>$+2km(I + k + 2 - \bar{x})(1 - \bar{x})$ | $n(mi / n + 2 - \bar{x})^2(m * i / n + 2)$<br>$+m(i + k + 2 - \bar{x})^2(i + k + 2)$<br>$+mk(1 - \bar{x})^2$ | $r < 0$ unless $m$<br>is large and $m > n$<br>$i = 2,\ k = 2$<br>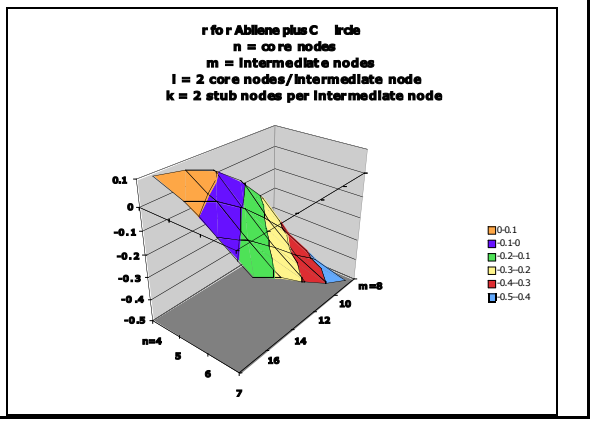 |

<table>
<tr>
<td>Poisson field of $L$ randomly oriented lines plus a circle</td>
<td>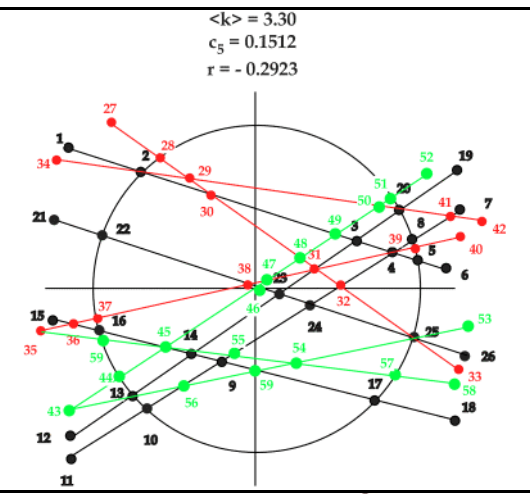
</td>
<td>$num_rows =$<br>$10L + \alpha L(L-1)/2$<br><br>$row_sum$<br>$= 32L$<br>$+16\alpha L(L-1)/2$</td>
<td>$2L(1-\bar{x})(4-\bar{x})$<br>$+2L(1-\bar{x})(4-\bar{x})$<br>$+6L(4-\bar{x})^2$<br>$+4\alpha L(4-\bar{x})^2(L-1)/2$</td>
<td>$2L(1-\bar{x})^2 + 2L(4-\bar{x})^2$<br>$+6L(4-\bar{x})^2$<br>$+4\alpha L(4-\bar{x})^2(L-1)/2$</td>
<td>$r < 0$<br>Intended to be a random version of HOT + circle<br>$\alpha \approx 0.5$ in simulations</td>
</tr>
<tr>
<td>Incompatibility networks [Zhang et al]</td>
<td>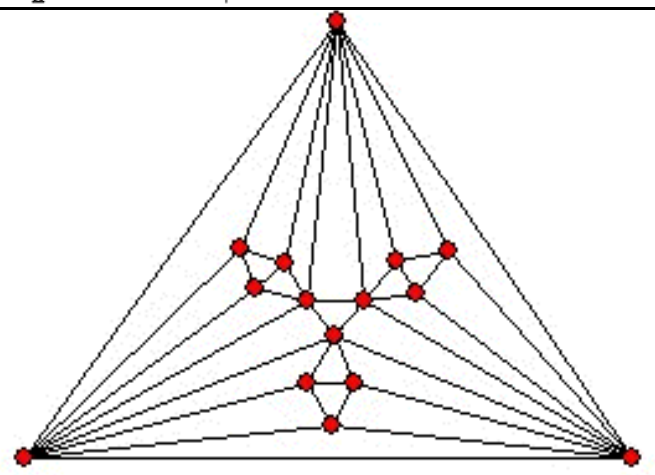</td>
<td>Network grows with time $t$ by building triangles inside existing triangles recursively</td>
<td>$$k_{nn}(k) = \frac{\left[8\left(\frac{4}{3}\right)^t - 4\left(\frac{2}{3}\right)^t\right](k-2)^{\ln 3/\ln 2}}{k(k-2)} + \frac{6}{k} + \left(1 - \frac{2}{k}\right)\frac{\ln(k-2)}{\ln 2} - 1.$$</td>
<td>Derivation of $k_{nn}$ is in [42]</td>
<td>$r < 0$ ($k_{nn}$ falls with $k$)<br>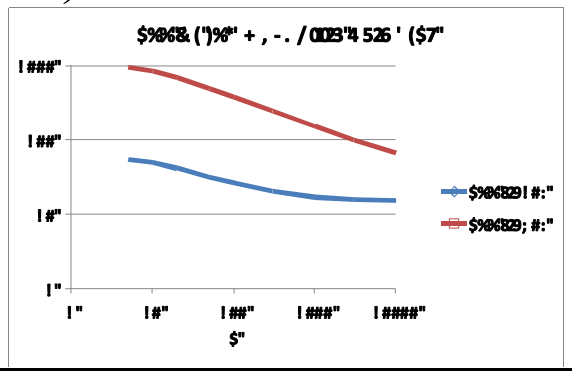</td>
</tr>
<tr>
<td>Planar unclustered graphs [Miralles et al]</td>
<td>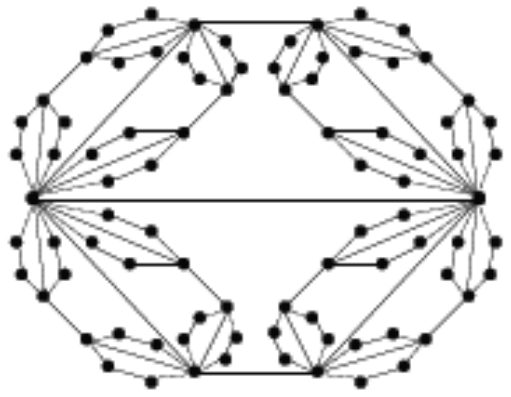</td>
<td>Network grows with time $t$ by adding new structures with multiplicity $d$ recursively</td>
<td>
<table>
<tr><th></th><th>$t=1$</th><th>$t=2$</th><th>$t=3$</th><th>$t=10$</th></tr>
<tr><td>$d=2$</td><td>$-0.1667$</td><td>$-0.0886$</td><td>$-0.0460$</td><td>$-0.0003$</td></tr>
<tr><td>$d=10$</td><td>$-0.4091$</td><td>$-0.2338$</td><td>$-0.1174$</td><td>$-0.0009$</td></tr>
<tr><td>$d=100$</td><td>$-0.4901$</td><td>$-0.2057$</td><td>$-0.0934$</td><td>$-0.0007$</td></tr>
</table>
$r$ for selected $t$ and $d$</td>
<td>Derivation of $r$ is in [43]</td>
<td>$r < 0$ approaching 0 from below as $t, d \to \infty$</td>
</tr>
</table>